\documentclass{LMCS}

\usepackage{enumerate}
\usepackage{hyperref}

\usepackage{amsmath}
\usepackage{amssymb}
\usepackage{graphicx}
\usepackage{url}
\usepackage{bbm}
\usepackage{diagrams}
\newcommand{\comment}[1]{}

\newif\ifproofs\proofstrue

\newcommand{\prefixclosure}[1]{#1\!\!\downarrow}

\newcommand{\eqdef}{=_{\mathrm{def}}}

\newcommand{\compunit}{*}

\newcommand{\forkn}{\mbox{\tt async}}

\newcommand{\fork}[1]{\forkn~#1}

\newcommand{\rforkn}{\mbox{\tt yield\_to}}
\newcommand{\rfork}[1]{\rforkn~#1}

\newcommand{\orm}[2]{#1~\mbox{\tt or}~#2}

\newcommand{\block}{\mbox{\tt block}}
\newcommand{\yield}{\mbox{\tt yield}}

\newcommand{\btrue}{\mbox{\tt true}}
\newcommand{\bfalse}{\mbox{\tt false}}
\newcommand{\unit}{\mbox{\tt skip}}
\newcommand{\assign}[2]{#1:=#2}
\newcommand{\Ifthenelse}[3]{\mbox{\tt if\ }#1 \mbox{\tt \ then\ } #2 \mbox{\tt \ else\ }#3}
\newcommand{\Whiledo}[2]{\mbox{\tt while\ }#1 \mbox{\tt \ do\ } #2}
\newcommand{\alt}{\,\,\vert\,\,}

\newcommand{\nt}[1]{{\langle #1 \rangle}}
\newcommand{\transrulens}[3]{#1 &\!\standards\!& #2 &\mbox{} \\}
\newcommand{\transrulenssimple}[2]{#1 &\standards& #2}

\newcommand{\Thread}{\mathrm{Pool}}
\newcommand{\PThread}{\mathrm{AProc}}
\newcommand{\Res}{\mathrm{Res}}
\newcommand{\Proc}{\mathrm{Proc}}
\newcommand{\type}{\!:\!}
\newcommand{\TSeq}{\mathrm{TSeq}}
\newcommand{\PSeq}{\mathrm{PSeq}}
\newcommand{\PPSeq}{\mathrm{PPSeq}}
\newcommand{\comp}{\circ}
\newcommand{\delay}{\mathrm{d}}
\newcommand{\async}{\mathrm{async}}

\newcommand{\donem}{\mathrm{Done}}
\newcommand{\finalm}{\mathrm{Ret}}
\newcommand{\final}{return}

\newcommand{\dsem}[1]{[\![{#1}]\!]}
\newcommand{\fst}[1]{\mathrm{fst}(#1)}
\newcommand{\clean}[1]{{#1}^c}
\newcommand{\shuffle}[2]{{#1}\bowtie{#2}}
\newcommand{\emptyseq}{\varepsilon}
\newcommand{\emptythread}{\varepsilon}
\newcommand{\prefix}{\leq_p}

\newcommand{\action}{\!\cdot\!}

\newcommand{\checkstore}[1]{\mathrm{check}(#1)}
\newcommand{\gotostore}[1]{\mathrm{goto}(#1)}

\newcommand{\mesh}[1]{\mathrm{mesh}(#1)}
\newcommand{\gofromtostore}[2]{(#1 \leadsto #2)}

\newcommand{\tgofromtostore}[3]{(#1 \leadsto #2 \leadsto #3)}

\newcommand{\onerun}[1]{\mathrm{run}(#1)}
\newcommand{\runs}[1]{\mathrm{runs}(#1)}
\newcommand{\observ}[1]{\mathrm{obs}(#1)}

\newcommand\fun[1]{\mathrm{#1\/}}
\newcommand{\Value}{\mathrm{Value\/}}
\newcommand{\Store}{\mathrm{Store\/}}
\newcommand{\BExp}{\mathrm{BExp\/}}
\newcommand{\NExp}{\mathrm{NExp\/}}
\newcommand{\Exp}{\mathrm{Com\/}}
\newcommand{\Vars}{\mathrm{Vars}}

\newcommand{\cmdapprox}{\preceq}

\newcommand{\update}{\mathrm{update}}
\newcommand{\lookup}{\mathrm{lookup}}
 \newcommand{\Cpo}{\omega\mathrm{Cpo}}
  
 \newcommand{\SL}{\omega\mathrm{SL}}
 \newcommand{\Pos}{\mathrm{Pos}}

\newcommand\standards{\mathbin{\longrightarrow}}
\newcommand\standardschoice{\mathbin{\longrightarrow_c}}
\newcommand\standardsaction{\mathbin{\longrightarrow_a}}
\newcommand{\avcxt}[1]{\avcxtsym[#1]}
\newcommand{\avcxtsym}{{\mathcal{E}}}
\newcommand{\ccxt}[1]{\ccxtsym[#1]}
\newcommand{\ccxtsym}{{\mathcal{C}}}

\newcommand{\hole}{[~]}

\newcommand{\mymod}[1]{|#1|}

\newcommand{\llaction}{\lhd}
\newcommand{\rlaction}{\rhd}

\newcommand{\rsync}{\mathrm{yield\_to}}

\newcommand{\Spawn}{\mathrm{Spawn}}

 \newcommand{\Trans}{\mathrm{Trans}}

 \newcommand{\BTrans}{\mathrm{BTrans}} 
 \newcommand{\halt}{\mathrm{halt}}
 
  \newcommand{\Id}{{\mathcal I}}

\newcommand{\markret}[1]{{#1}^m}
\newcommand{\finishn}{\mbox{\tt finish}}
\newcommand{\finish}[1]{\finishn~#1}
\newcommand{\infix}{\mathrel}

\newcommand{\parop}{\infix{\mid\mid}}

\def\doi{6 (4:2) 2010}
\lmcsheading%
{\doi}
{1--39}
{}
{}
{Sep.~17, 2010}
{Oct.~20, 2010}
{}   

\begin{document}

\title[A Model of Cooperative Threads]{A Model of Cooperative Threads\rsuper*}

\author[M.~Abadi]{Mart\'\i n Abadi\rsuper a}	
\address{{\lsuper a}Microsoft Research, Silicon Valley; University of California, Santa Cruz}	
\email{abadi@microsoft.com}  

\author[G.~D.~Plotkin]{Gordon D.\ Plotkin\rsuper b}	
\address{{\lsuper b}Microsoft Research, Silicon Valley; LFCS, University of Edinburgh}	
\email{gdp@inf.ed.ac.uk}  



\titlecomment{{\lsuper*}A conference version of this paper has appeared as~\cite{AP09}}




\begin{abstract}
\noindent We develop a model of concurrent imperative programming with
threads. We focus on a small imperative language with cooperative
threads which execute without interruption until they terminate or
explicitly yield control.  We define and study a trace-based
denotational semantics for this language; this semantics is fully
abstract but mathematically elementary.  We also give an equational
theory for the computational effects that underlie the language,
including thread spawning.  We then analyze threads in terms of the
free algebra monad for this theory.
\end{abstract}

\date{}
\maketitle

\section{Introduction}
\label{s:intro}

In the realm of sequential programming, 
semantics, whether operational or denotational, provides a
rich understanding of programming constructs and languages, and 
serves a  broad range of purposes.
These include, for instance, the study of verification techniques and
the reconciliation of effects with functional programming
via monads.
With notorious difficulties, these two styles of semantics have been
explored for concurrent programming, and, by now, a substantial body of
work provides various semantic accounts of concurrency.  Typically,
that work develops semantics for 
languages with parallel-composition constructs
and various communication mechanisms.

Surprisingly, however, that work provides only a limited understanding
of threads. 
It includes several operational semantics of languages
with threads, sometimes with operational notions of equivalence, 
e.g.,~\cite{BMT,essenceofcml,jeffreycore,Jeffrey:FAM};
denotational semantics of those languages seem to be much rarer,
and to address message passing rather than shared-memory 
concurrency, e.g.,~\cite{DBLP:journals/tcs/FerreiraH99, LICS::Jeffrey1995}.
Yet threads are in widespread use,
often in the context of elaborate 
shared-memory systems and languages for
which a clear semantics would be beneficial.

In this paper, we investigate a model of concurrent imperative
programming with threads.  We focus on \emph{cooperative} threads which
execute, without interruption, until they either terminate or else explicitly
yield control.
Non-cooperative threads, that is, threads with preemptive scheduling, can be
seen as threads that yield control at every step. In this sense, they are a
special case of the cooperative threads that we study. 

Cooperative threads appear in several systems, programming models, and
languages.  Often without much linguistic support, they have a long
history in operating systems and databases, e.g.,~\cite{SQL}.  
Cooperative threads also arise in other
contexts, such as Internet services and synchronous
programming~\cite{adya,capriccio,Boussinot06,Boudol,amadio:resource}.  Most recently,
cooperative threads are central in two models for programming with
transactions, Automatic Mutual Exclusion (AME) and Transactions with
Isolation and Cooperation
(TIC)~\cite{isard:automatic,smaragdakis:transactions}.
AME is one of the main starting points for our research.  The intended
implementations of AME rely on software transactional
memory~\cite{shavit:software} for executing multiple cooperative
threads simultaneously. However, concurrent transactions do not appear
in the high-level operational semantics of the AME constructs~\cite{abadi:semantics}. 
Thus,
cooperative threads and their semantics are of interest independently
of the details of possible transactional implementations.

We define and study three semantics for an
imperative language with primitives for spawning threads, yielding
control, and blocking execution.
\begin{enumerate}[$\bullet$]

\item We obtain an operational semantics by a straightforward
adaptation of previous work. 
In this semantics, we describe the meaning of
a whole program in terms of small-step transitions between states
 in which spawned threads are kept in a thread pool. 
This semantics serves
as a 
reference point.

\item We also define a more challenging compositional denotational semantics. 
The meaning of a command is
a prefix-closed set of traces. 
Prefix-closure 
arises because we are primarily interested in safety properties,
that is, in ``may'' semantics. 
Each trace is roughly a sequence
of transitions, where each transition is a pair of stores, and a store
is a mapping from variables to values. 
We establish adequacy and full-abstraction 
theorems with respect to the operational semantics.
These results require
several non-trivial choices in the definition of the denotational semantics.

\item Finally, we define a semantics based on the algebraic theory of
effects. More precisely, we give an equational theory for the computational effects that underlie the language, and
analyze threads in terms of the free algebra monad for this theory.
This definition is more principled and systematic; it
explains threads with standard semantic structures, in the context of
functional programming.
As we show, furthermore, we obtain our denotational semantics as a special case.
\end{enumerate}

\noindent Section~\ref{sec:calculus} introduces our language and
Section~\ref{sec:osemantics} defines its operational semantics.
Section~\ref{sec:dsemantics} develops 
its denotational semantics.
Section~\ref{sec:AdequacyFull} presents our adequacy 
 and full-abstraction theorems (Theorems~\ref{thm:adequacy} and~\ref{thm:fullabs}).
Section~\ref{sec:algebra} concerns the algebraic \hbox{theory} of effects
and the analysis of the denotational semantics in this monadic setting
(Theorem~\ref{thm:view}). Section~\ref{sec:conclusion} concludes.

\section{The Language}
\label{sec:calculus}

Our language is an extension of a basic imperative language with
assignments, sequencing, conditionals, and while loops
(IMP~\cite{Winskel93}). Programs are written in terms of a finite set of
variables $\Vars$, whose values are natural numbers. 
In addition to those standard constructs, our language includes:
\begin{enumerate}[$\bullet$] 
\item A
construct for executing a command in an asynchronous thread. 
Informally, $\fork{C}$
forks off the execution of $C$. This execution is asynchronous, and
will not happen if the present thread keeps running without ever
yielding control, or if the present thread blocks without first yielding control. 

\item A construct for yielding control. Informally, $\yield$ indicates
that any pending thread may execute next, as may the current thread.

\item  A construct for blocking.
Informally, $\block$ halts the execution of the 
entire program, 
even if there are pending threads that could otherwise make progress.
\end{enumerate}

\noindent We define the syntax of the language in
Figure~\ref{fig:language}.  We do not detail the constructs on
numerical and boolean expressions, which are as usual.

\begin{figure}
{\noindent\hrulefill\par}
\[
\begin{array}[t]{rclrll}
b &\in& \BExp &=& \ldots \\
e &\in& \NExp &=& \ldots \\
C,D &\in& \Exp &=& \unit \\
        &&&\alt& \assign{x}{e} \qquad (x \in \Vars)\\
        &&&\alt& {C} ; {D} \\
        &&&\alt& \Ifthenelse{b}{C}{D}\\
        &&&\alt& \Whiledo{b}{C}\\
        &&&\alt& \fork{C} \\
        &&&\alt& \yield \\
        &&&\alt& \block
\end{array}
\]
\caption{\label{fig:language}
         Syntax.}
{\noindent\hrulefill\par}
\end{figure}

Figure~\ref{fig:excode} gives an illustrative example. It shows
a piece of code that 
spawns the asynchronous execution of $\assign{x}{0}$,
then executes  $\assign{x}{1}$ and yields, then resumes but 
blocks unless the predicate $x=0$
holds, then executes $\assign{x}{2}$. 
%
\begin{figure}[h]
{\noindent\hrulefill\par}
\[
\begin{array}{l}
\fork{\assign{x}{0}} ;\\
\assign{x}{1};\\
\yield ;\\
\Ifthenelse{x=0}{\unit}{\block};\\
\assign{x}{2}
\end{array}
\]
\caption{\label{fig:excode}
         Example command.}
{\noindent\hrulefill\par}
\end{figure}
%
%
The execution of $\assign{x}{0}$ may happen once the $\yield$ statement is reached.
With respect to safety properties, the conditional blocking amounts to
waiting for $x=0$ to hold. 
More generally, AME's 
${\mbox{\tt blockUntil\ }b}$ can be written
$\Ifthenelse{b}{\unit}{\block}$.

More elaborate uses of blocking are
possible too, and supported by lower-level semantics and actual transactional
implementations~\cite{isard:automatic,abadi:semantics}.  
In those implementations,
blocking may cause a roll-back and a later retry at an appropriate
time.  We regard roll-back as an interesting aspect of some possible
implementations, but not as part of the high-level semantics of our
language, which is the subject of this work.

Thus, our language is basically a fragment of the AME calculus~\cite{abadi:semantics}. It omits
higher-order functions and references.
It also omits ``unprotected sections'' for
non-cooper\-ative code, particularly legacy code. Non-cooperative code
can however be modeled as code with pervasive calls to $\yield$
(at least with respect to the simple, strong memory models that we use throughout 
this paper; cf.~\cite{GMP06}).
See Section~\ref{sec:conclusion} for further discussion of possible 
extensions to our language.


\section{Operational Semantics}
\label{sec:osemantics}

We give an operational semantics for our language. Despite some
subtleties, this semantics is not meant to be challenging.  It 
is given in terms of 
small-step transitions between states. Accordingly,
we define states, evaluation contexts, and the transition relation.

%

\subsection{States}\label{sec:states}

%
\begin{figure}[t]
{\noindent\hrulefill\par}
\[
\begin{array}{rcrcll}
\Gamma &\in& \fun{State} 
                &=& {\Store}\times \fun{ComSeq} \times \fun{Com} \\
\sigma &\in& {\Store} 
                        &=& \Vars \rightarrow \Value \\
n &\in& \Value &=& \mathbb{N}\\
T &\in& \fun{ComSeq} &=& \fun{Com}^* 
\end{array}
\]
\caption{State space.
         \label{fig:machine:A}}
{\noindent\hrulefill\par}
\end{figure}

As described in Figure~\ref{fig:machine:A}, a 
\emph{state} $\Gamma = \nt{\sigma, T, C}$ consists of the following components: 
\begin{enumerate}[$\bullet$]
\item a \emph{store} $\sigma$ which is  a mapping of the given finite set 
$\Vars$ of variables to a set $\Value$ of values, which we take to be the set of natural numbers;
\item a finite sequence of commands $T$ which we call the \emph{thread pool};
\item a distinguished \emph{active} command $C$.
\end{enumerate}
We write
$\sigma[x\mapsto n]$ for the store that agrees with $\sigma$ except
at $x$, which is mapped to $n$.
We write $\sigma(b)$ for the boolean denoted by $b$ in $\sigma$,
and $\sigma(e)$ for the natural number denoted by $e$ 
in $\sigma$, similarly.  
We write $T.T'$ for the concatenation of two thread pools $T$ and $T'$.

\subsection{Evaluation Contexts}

As usual, a context is an expression with a hole $\hole$,
and an evaluation context is a context of a particular kind.
Given a context $\ccxtsym$ and an expression $C$, we write
$\ccxt{C}$ for the result of placing $C$ in the hole in $\ccxtsym$.
We use the evaluation contexts defined by the grammar:
\[
\begin{array}{rcl}
\avcxtsym 
  &=& \hole\alt {\avcxtsym}; C %
\end{array}
\]

\subsection{Steps}

\begin{figure*}[t]
{\noindent\hrulefill\par}
\[
\begin{array}{lcl@{\quad}l}
\transrulens
        {\nt{\sigma,T,\avcxt{\assign{x}{e}}}}
        {\nt{\sigma[x\mapsto n],T,\avcxt{\unit}}\qquad \quad \; \mbox{(\rm if \ }\sigma(e) = n)}
        {}
\\ 
\transrulens
        {\nt{\sigma,T,\avcxt{\unit ; C}}}
        {\nt{\sigma,T,\avcxt{C}}}
        {}
\\ 
\transrulens
        {\nt{\sigma,T,\avcxt{\Ifthenelse{\!b\!}{\!C\!}{\!D}}}}
        {\nt{\sigma,T,\avcxt{C}}\qquad \qquad \qquad\qquad\mbox{(\rm if \ }\sigma(b) = \btrue)}        
        {}
\\ 
\transrulens
        {\nt{\sigma,T,\avcxt{\Ifthenelse{\!b\!}{\!C\!}{\!D}}}}
        {\nt{\sigma,T,\avcxt{D}}        \qquad \qquad \qquad\qquad \mbox{(\rm if \ }\sigma(b) = \bfalse)}
        {}
\\ 
\transrulens
        {\nt{\sigma,T,\avcxt{\Whiledo{\!b\!}{\!C}}}}
        {\nt{\sigma,T,\avcxt{\Ifthenelse{\!b\!}{\!\!(C ; \Whiledo{\!b\!}{\!C})\!\!}{\!\unit}}}}
        {}
\\ 
\transrulens
        {\nt{\sigma,T,\avcxt{\fork{C}}}}
        {\nt{\sigma,T.C,\avcxt{{\unit}}}}
        {}
\\ 
\transrulens
        {\nt{\sigma,T,\avcxt{\yield}}}
        {\nt{\sigma,T.\avcxt{\unit},\unit}}
        {}
\\ 
\transrulens
        {\nt{\sigma,T.C.T',\unit}}
        {\nt{\sigma,T.T',C}}
        {}
\end{array}
\]
\caption{Transition rules of the abstract machine.
         \label{fig:machinestrong}}
{\noindent\hrulefill\par}
\end{figure*}

A transition $\Gamma \standards \Gamma'$ takes an execution
from one state to the next.
Figure~\ref{fig:machinestrong} 
gives rules that specify the transition relation.
According to these rules,
when the active command is $\unit$, a command
from the pool becomes the active command.
It is then evaluated as such 
until it produces $\unit$, yields, or blocks.
No other computation is interleaved with this evaluation.
Each evaluation step produces a new state,
determined by 
decomposing the active command into an evaluation context and a
subexpression that describes a computation step (for instance, 
a yield or a conditional).
In all cases at most one rule applies.
In two cases, no rule applies. The
first is when the active command is $\unit$ and the pool is empty;
this situation corresponds to \emph{normal} termination.  The second is
when the active command is \emph{blocked}, in the sense that it has the
form $\avcxt{\block}$; this situation is an  \emph{abnormal} 
termination.

We write $\Gamma \standardschoice\Gamma'$ when
$\Gamma \standards\Gamma'$ via the last rule, 
and call this 
a \emph{choice} transition.
We write $\Gamma \standardsaction \Gamma'$ when
$\Gamma \standards\Gamma'$ via the other rules, and call this 
an \emph{active} transition.
Active transitions are deterministic, i.e., 
if $\Gamma \standardsaction \Gamma'$ and
$\Gamma \standardsaction \Gamma''$ then $\Gamma' = \Gamma''$.

\section{Denotational Semantics}
\label{sec:dsemantics}

Next we give a compositional denotational semantics for the same
language.  Here, the meaning of a command is a prefix-closed set of
traces, where each trace is roughly a sequence of transitions, and
each transition is a pair of stores.

The use of sequences of transitions goes back at least to Abrahamson's
work~\cite{abrahamson:modal} and appears in various studies of 
parallel composition~\cite{TCS::AbadiP1993,horita,IC::Brookes1996,Brookes02}.  
However, the treatment of threads requires some
new non-trivial choices. For instance, transition sequences, as we
define them, include markers to indicate not only normal termination but also
the return of the main thread of control. Moreover, although these
markers are similar, they are attached to traces in different ways,
one inside pairs of stores, the other not. Such details are crucial
for adequacy and full abstraction.

Also crucial to full abstraction is minimizing the information that
the semantics records. More explicit semantics will typically be more
transparent, for instance, in detailing that a particular step in a computation causes the spawning of a thread, but
will
consequently fail to be
fully abstract.

Section~\ref{sec:informal} is an informal introduction to some of the
details of the semantics.  Section~\ref{sec:trans} defines transition
sequences and establishes some 
notation.
Sections~\ref{sec:semcommand} and~\ref{sec:pool} define the
interpretations of commands and thread pools, respectively.
Section~\ref{sec:eq} discusses semantic
equivalences.

\subsection{Informal Introduction}
\label{sec:informal}

As indicated above, the meaning of a command will be a prefix-closed set of
traces, where each trace is roughly a sequence of transitions, and
each transition is a pair of stores. 
Safety properties---which pertain to what ``may'' happen---are 
closed under prefixing, hence the prefix-closure condition.
Intuitively, when the meaning of a command includes
a trace $(\sigma_1,\sigma_1') (\sigma_2,\sigma_2') \ldots$,
we intend that the command may start executing with store $\sigma_1$, 
transform it to $\sigma_1'$, yield, then resume with store $\sigma_2$, 
transform it to $\sigma_2'$, yield again, and so on.

In particular, the meaning of $\block$ will consist of
the empty sequence $\emptyseq$.  The
meaning of $\yield ; \block$ will consist of the empty
sequence $\emptyseq$ plus every sequence of the form $(\sigma,
\sigma)$, where $\sigma$ is any store. Here, the pair 
$(\sigma, \sigma)$ is a ``stutter'' that represents immediate yielding.

If the meaning of a command $C$ includes $(\sigma_1,\sigma_1') \ldots
(\sigma_n,\sigma_n') $ and the meaning of a command $D$ includes
$(\sigma_n',\sigma_n'') \ldots (\sigma_m,\sigma_m') $, one might
naively expect that the meaning of $C ; D$ would contain
$(\sigma_1,\sigma_1') \ldots (\sigma_n,\sigma_n'') \ldots
(\sigma_m,\sigma_m') $, which is obtained by concatenation plus a
simple local composition between $(\sigma_n,\sigma_n')$ and
$(\sigma_n',\sigma_n'')$. Unfortunately, this naive expectation is
incorrect.
In a trace $(\sigma_1,\sigma_1') (\sigma_2,\sigma_2') \ldots$, some of
the pairs may represent steps taken by commands to be executed
asynchronously. Those steps need not take place before any further
command $D$ starts to execute. 

Accordingly, computing the meaning of $C ; D$
requires shuffling suffixes of traces in $C$ with traces in $D$. The
shuffling represents the interleaving of $C$'s asynchronous work with
$D$'s work. We introduce a special {\final}
marker 
``$\finalm$'' in order to indicate
how the traces in $C$ should be parsed for this composition.
In particular, when $C$ is of the form $C_1 ; \fork{(C_2)}$, any 
occurrence of ``$\finalm$'' in the meaning of $C_2$ will not appear
in the meaning of $C$.
The application
of $\forkn$ erases any occurrence 
of ``$\finalm$'' from the meaning of $C_2$---intuitively, because $C_2$ does not
return control to its sequential context.

For example, 
the meaning of the command
\[\assign{x}{n} ; \yield ; \assign{x}{n'}
\]
will contain the trace
\[(\sigma, \sigma[x\mapsto n])(\sigma',\sigma'[x\mapsto n'] \ \finalm)
\]
for every $\sigma$ and $\sigma'$.
On the other hand, 
the meaning of the command
\[\assign{x}{n} ; \fork{(\assign{x}{n'})} ; \yield 
\]
will contain the trace
\[
(\sigma, \sigma[x\mapsto n] \ \finalm)(\sigma',\sigma'[x\mapsto n'])
\]
for every $\sigma$ and $\sigma'$. The different positions of the
marker $\finalm$ correspond to different junction points for any commands
to be executed next. 

If the meaning of $C$ contains $u (\sigma_n,\sigma_n' \ \finalm) u'$
and the meaning of $D$ contains  $(\sigma_n',\sigma_n'') v$,
then the meaning of $C ; D$ contains $u (\sigma_n,\sigma_n'') w$,
where $w$ is a shuffle of $u'$ and $v$. 
Notice that the marker from $u (\sigma_n,\sigma_n' \ \finalm) u'$ disappears
in this combination. The marker in $u (\sigma_n,\sigma_n'') w$, if present,
comes from $(\sigma_n',\sigma_n'') v$.
An analogous combination applies when 
the meaning of $C$ contains $u (\sigma_n,\sigma_n' \ \finalm) u'$
and the meaning of $D$ contains  $(\sigma_n',\sigma_n'' \ \finalm) v$ (a trace
that starts with a transition with a marker).
Moreover, if 
the meaning of $C$ contains a trace without any occurrence of the 
marker $\finalm$,
then this trace is also in the meaning of $C ; D$: the absence of
a marker makes it impossible to combine this trace with traces from $D$.

An additional marker, ``$\donem$'', 
ends traces that represent complete
normally terminating executions. Thus, the meaning of $\unit$
will consist of the empty
sequence $\emptyseq$ and every sequence of the form $(\sigma,
\sigma \ \finalm)$
plus every sequence of the form $(\sigma,
\sigma \ \finalm) \donem$.
Contrast this with the meaning of 
$\yield ; \block$ given above.

It is possible for a trace to contain a 
$\finalm$ marker but not a
$\donem$ marker. Thus, the meaning of $\fork{(\block)}$ will contain the
empty sequence $\emptyseq$ plus every sequence of the form $(\sigma,
\sigma \ \finalm)$, but not $(\sigma, \sigma \ \finalm) \donem$. 

More elaborately, the meaning of the code of Figure~\ref{fig:excode} will contain
all traces of the form
\[
(\sigma, \sigma[1])
(\sigma[1], \sigma[0])
(\sigma[0], \sigma[2] \ \finalm) \donem
\]
where we write $\sigma[n]$ as an abbreviation for $\sigma[x \mapsto n]$.
These traces model normal termination after 
taking the $\btrue$ branch of the conditional
$\Ifthenelse{x=0}{\assign{x}{2}}{\block}$.
The meaning will also contain all prefixes of those traces, which model
partial executions---including those that take
the $\bfalse$ branch of the conditional and terminate abnormally.

The two markers are somewhat similar. However,
note that 
$(\sigma,\sigma'\ \finalm)$ 
is a prefix of 
$(\sigma,\sigma'\ \finalm)\donem$,
but
$(\sigma,\sigma')$ 
is not a prefix of
$(\sigma,\sigma'\ \finalm)$.
Such differences are essential.

\subsection{Transitions and Transition Sequences}
\label{sec:trans}

A \emph{plain transition} is a pair of stores  $(\sigma,\sigma')$.
A \emph{{\final}   transition} is a pair of stores $(\sigma,\sigma' \ \finalm)$ in which the second is adorned
with the marker $ \finalm$.
A \emph{transition} is a plain transition or a {\final}  transition.

A \emph{main-thread transition sequence} (hereunder simply: {\em transition sequence}) is a finite (possibly empty) sequence, beginning with  a sequence of transitions, of which at most one (not necessarily the last) is a {\final} transition, and optionally followed by the marker $\donem$ if one of the transitions is a {\final} transition.
 We write $\TSeq$ for the set of transition sequences. 

A \emph{pure transition sequence} is a finite sequence of plain transitions, possibly followed by a marker
 $\donem$. Note that such a  sequence need not be a transition sequence. 
It is \emph{proper} if it is not equal to $\donem$. We write $\PSeq$ for the set of pure transition sequences, and $\PPSeq$ for the subset of the proper ones.

We use the following notation:
\begin{enumerate}[$\bullet$]

\item 
We typically let $u$, $v$, and $w$ range over transition sequences or pure transition sequences, and let $t$ range over non-empty ones.

\item We write $u \prefix v$ for the prefix relation between sequences $u$ and $v$ (for both kinds of sequences, pure or not).  
For example, as mentioned above, we have that 
$(\sigma,\sigma'\ \finalm) \prefix  (\sigma,\sigma'\ \finalm)\donem$,
but
$(\sigma,\sigma') {\not\prefix} (\sigma,\sigma'\ \finalm)$.

\item A set $P$ is \emph{prefix-closed} if whenever  $u \prefix v \in P$  then $u \in P$. 
We write $\prefixclosure{P}$ for the least prefix-closed 
set that contains $P$.

\item 
For a non-empty sequence of transitions $t$, we write $\fst{t}$ for the first
store of the first transition of $t$.
\item 
For a transition sequence $u$, we write $\clean{u}$ for the pure transition sequence obtained by {\em cleaning $u$}, which means removing the $ \finalm$ marker, if present, from~$u$.
\item We let $\tau$ range over stores and stores with {\final}  markers.

\end{enumerate}

\subsection{Interpretation of Commands}
\label{sec:semcommand}

\subsubsection*{Preliminaries}

We let $\Proc_{}$ be the collection of the non-empty prefix-closed sets of  transition sequences, and let $\Thread_{}$ be the collection of the non-empty prefix-closed sets of pure transition sequences. 
Under the subset partial ordering, $\Proc_{}$ and $\Thread_{}$ are both
$\omega$-cpos (i.e., partial orders with sups of increasing sequences) with least element $\{\emptyseq\}$. 
We interpret commands  as elements of $\Proc_{}$. We use $\Thread_{}$ as an auxiliary 
$\omega$-cpo; below it also serves for the semantics of thread pools.
We also let $\PThread_{}$ be the sub-$\omega$-cpo of $\Thread_{}$ of all non-empty prefix-closed sets of proper pure transition sequences. 
We think of such sets as modeling asynchronous threads, spawned by an active thread; the difference from $\Thread$ is that the latter also contains an element 
that models the empty thread pool.

We define a continuous cleaning function 
\[\clean{-}\type \Proc_{} \rightarrow \PThread_{}\] 
by:
\[\clean{P} = \{\clean{u} \mid u \in P\}\]
(Continuous functions are those preserving all sups of increasing sequences.)

We define the set $\shuffle{u}{v}$  of \emph{shuffles} of a pure transition sequence $u$ with a sequence $v$, whether a transition sequence or a pure transition sequence,  as follows:
\begin{enumerate}[$\bullet$]

\item  If neither finishes with  $\donem$,
  their set of shuffles is defined as usual for finite sequences.

\item If $u$ does not finish with $\donem$, then a shuffle of $u$ and
  $v\,\donem$ is a shuffle of $u$ and $v$. Similarly, if
  $v$ does not finish with $\donem$, then a shuffle of $u\,\donem$
  and $v$ is a shuffle of $u$ and $v$. 

\item A shuffle of  $u\,\donem$ and $v\,\donem$
  is a shuffle of $u$ and $v$   followed by $\donem$.
\end{enumerate}
If both $u$ and $v$ are pure transition sequences then so is every element of $u \bowtie v$; if $u$ is a pure transition sequence and $v$ is a transition sequence, then every element of $u \bowtie v$ is a transition sequence.  
\begin{lem} \label{shuffle-assoc}
 For any $u$,$v$, and $w$ where either:
 \begin{enumerate}[$\bullet$]
 \item all three are pure transition sequences, or
  \item $u$ and $v$ are pure transition sequences, and $w$ is a transition sequence
 \end{enumerate}
  we have:
\[\bigcup \{ v'  \bowtie w \mid v' \in u \bowtie v\} =  \bigcup \{u \bowtie v'\mid v' \in v \bowtie w\}\eqno{\qEd}\]
\end{lem}
We define a continuous \emph{composition} function 
\[\comp: \Proc_{}^2 \rightarrow \Proc_{}
\] 
by:
\[\begin{array}{rcl} P\comp Q & =  & \{ u (\sigma, \tau) v \mid
               \exists  \sigma', w, w'.\;  u (\sigma, \sigma'\  \finalm) w \in P, \\
&&               \quad 
               (\sigma', \tau) w' \in Q,   v \in \shuffle{w}{w'} \}\\
 & &                 \cup\ \{ u \mid u \in P  \mbox{\rm \ with no {\final}  transition}\}
 \end{array}\]
Composition is associative with two-sided unit, given by:
\[\begin{array}{rcl} 
\compunit & = & \prefixclosure{\{ (\sigma, \sigma \  \finalm) \donem \mid \sigma \in {\Store}\}}
\end{array}\]

We also define a continuous \emph{delay} function 
\[
\delay: \Proc_{} \rightarrow \Proc_{}
\] 
by:
\[\begin{array}{rcl} 
\delay(P) & = & \prefixclosure{
\{(\sigma, \sigma) u \mid \sigma \in {\Store}, u \in P\}}
\end{array}\]
Thus, $\delay(P)$ is
 $P$ preceded by  
all possible stutters (plus $\varepsilon$).
Similarly, we define a continuous  function
\[
\async: \PThread_{} \rightarrow \Proc_{}
\] 
by:
\[\begin{array}{rcl} 
\async(Q) & = & \prefixclosure{
\{(\sigma, \sigma\ \finalm) u \mid \sigma \in {\Store}, u \in Q\}}
\end{array}\]
Thus, for $P \in \Proc_{}$, 
$\async(\clean{P})$ differs from $\delay(P)$ only in the placement of
the marker $\finalm$.

\subsubsection{Interpretation}

The denotational semantics 
\[\dsem{\cdot}\type \Exp \longrightarrow \Proc_{}\] 
maps a command to a non-empty prefix-closed set of transition
sequences.  
We define it in Figure~\ref{fig:den}.
There, the interpretation of loops relies on the following approximations:
\[
\begin{array}{lcl}
(\Whiledo{b}{C})_0 & = & \block\\
(\Whiledo{b}{C})_{i+1} & = & \Ifthenelse{b}{(C ; (\Whiledo{b}{C})_i)}{\unit}
\end{array}
\]
The 0-th approximant corresponds to divergence, which here we identify with blocking.

\begin{figure*}[t]
{\noindent\hrulefill\par}
\[
\begin{array}{rcl}
\dsem{\unit} & = & \compunit\\

\dsem{\assign{x}{e}} & = & \prefixclosure{\{ (\sigma, \sigma[x\mapsto n] \  \finalm)\donem \mid  \sigma \in {\Store}, \sigma(e) = n\}}\\

\dsem{ C ; D } & = &  \dsem{C}\comp \dsem{D}\\

\dsem{\Ifthenelse{b}{C}{D}} & = & 
\prefixclosure{\{ t \mid t \in \dsem{C}, \mbox{\rm non-empty}, \fst{t}(b) = \btrue\}} \\
 && 
\cup
\prefixclosure{\{ t \mid t \in \dsem{D}, \mbox{\rm non-empty}, \fst{t}(b) = \bfalse\}}\\

\dsem{\Whiledo{b}{C}} & = & \cup_i \dsem{(\Whiledo{b}{C})_i}\\

\dsem{\fork{C}} & = &  \async(\clean{\dsem{C}})\\

\dsem{\yield} & = & \delay(\compunit)\\

\dsem{\block} & = & \{\emptyseq\}

\end{array}
\]
\caption{Denotational semantics.
         \label{fig:den}}
{\noindent\hrulefill\par}
\end{figure*}

We straightforwardly extend the semantics
to contexts, so that
$$\dsem{\ccxtsym}: \Proc_{} \rightarrow \Proc_{}$$ is a continuous function on $\Proc_{}$. This function is defined
by induction on the form of $\ccxtsym$, with the usual clauses of the definition
of $\dsem{\cdot}$ plus $\dsem{\hole}(P) = P$.

\begin{prop}\label{prop:comp}
$\dsem{\ccxt{C}}  = \dsem{\ccxtsym}(\dsem{C})$.
Therefore, if $\dsem{C} \subseteq \dsem{D}$ then $\dsem{\ccxt{C}}  \subseteq \dsem{\ccxt{D}}$.
\qed
\end{prop}

\subsection{Interpretation of Thread Pools}
\label{sec:pool}

As an auxiliary definition, it is important to have also an
interpretation of thread pools as elements of $\Thread_{}$. We develop one in this section.

\subsubsection{Preliminaries}

We define a continuous shuffle operation  \[\bowtie\; \type (\Thread_{})^2 \rightarrow  \Thread_{}\]
 at this level by:
  \[P \bowtie Q = \bigcup_{u \in P, v \in Q} u \bowtie v \]
The shuffle operation is commutative and associative, with unit  $I \eqdef \{\emptyseq, \donem\}$;
associativity follows from Lemma~\ref{shuffle-assoc}. 

We define the set of right shuffles  $u\rlaction v$ of a pure transition sequence $u$ with a transition sequence $v$ by setting 
\[
u\rlaction (\sigma, \tau) v = \{(\sigma, \tau)w \mid w \in u \bowtie {v}\}
\] and 
\[
u \rlaction \emptyseq = \{\emptyseq\}
\]
\comment{GDP: the following version is wrong, I think:
$u\action v = \{v\}$ when there is no $ \finalm$ marker in $v$ and setting 
$u\action (v(\sigma, \sigma' \ \finalm) v') =  \{ v(\sigma, \sigma' \ \finalm) w \mid w \in u \bowtie v\} \cup \{\emptyseq\} $.}
We then define 
\[\async\type \Thread_{} \times \Proc_{} \longrightarrow \Proc_{}
\] 
by:
 \[\async(P,Q) = \bigcup_{u \in P, v \in Q} u \rlaction v\]
The use of the notation $\async$ for both a unary and a binary operation
is a slight abuse, though in line with the algebraic theory of effects: see the discussion in
Section~\ref{sec:algebra}. In this regard note the equality $\async(P) \comp Q = \async(P,Q)$
(and the equality $\dsem{\yield}\comp P = \delay(P)$ points to the corresponding relationship between $\delay$ and $\dsem{\yield}$).

\subsubsection{Interpretation}

We define the semantics of thread pools by:
\[\dsem{C_1, \ldots, C_{n}} = \clean{\dsem{C_1}}\bowtie \ldots \bowtie \clean{\dsem{C_{n}}} \quad \mbox{     ($n \geq 0$)}\]
intending that $\dsem{\emptythread} = I$. 
For any thread pool $T$, 
	$\donem \in \dsem{T}$ iff $T = \emptythread$
(because, for all $C$, 
$\donem \notin \dsem{C}^{c}$ and, for all $P$ and $Q$, 
$I \subseteq P \bowtie Q$ iff $I \subseteq P$ and $I \subseteq Q$).
Further, we set $\dsem{T,C} = \async(\dsem{T},\dsem{C})$.

\begin{lem} For all $P, Q \in \Thread_{}$ and $R \in \Proc_{}$ we have:
\begin{enumerate}[(1)]
	\item $\async(P\bowtie Q, R) = \async(P, \async(Q,R))$
	\item $\async(I,R) = R$
\end{enumerate}
\end{lem}
\proof For the first part, one shows for all pure transition sequences $u$ and $v$ and transition sequences $w$ that:
\[\bigcup \{ v'  \rlaction w \mid v' \in u \bowtie v\} =  \bigcup \{u \rlaction v'\mid v' \in v \rlaction w\}\]
To this end, one proceeds by cases on $w$, using Lemma~\ref{shuffle-assoc}. The second part is obvious.
\qed
\subsection{Equivalences}
\label{sec:eq}

An attractive application of denotational semantics is in proving
equivalences and implementation relations between commands. 
Such denotational 
proofs tend to be simple calculations.
Via adequacy and full-abstraction
results (of the kind established in Section~\ref{sec:AdequacyFull}), 
one then obtains operational results that would typically be much harder to obtain
directly by operational arguments.

As an example, we note that
we have the following equivalence:
\[
\dsem{\fork{(C ; \yield ; D)}}
= 
\dsem{(\fork{(C; \fork{(D)})}}
\]
This equivalence follows from three facts:
\begin{enumerate}[$\bullet$]
\item  We have:  \[\begin{array}{lcl}
\clean{\dsem{\yield ; D}} & =  & \clean{\dsem{\fork{(D)}}}\\
&  = &
\prefixclosure{\{ (\sigma, \sigma) \clean{u} \mid \sigma \in {\Store}, u \in \dsem{D}\}};
\end{array}\]
\item whenever $\clean{\dsem{D_1}} = \clean{\dsem{D_2}}$,
$\clean{\dsem{C ; D_1}} = \clean{\dsem{C; D_2}}$;
\item whenever $\clean{\dsem{D_1}} = \clean{\dsem{D_2}}$,
$\dsem{\fork{(D_1)}} = \dsem{\fork{(D_2)}}$.
\end{enumerate}
This particular equivalence is interesting for two reasons:
\begin{enumerate}[$\bullet$]

\item It models an implementation strategy (in use in AME)
where, when executing $C ; \yield ; D$, 
the
$\yield$ causes a new asynchronous thread for $D$ to be added to the thread
pool.

\item It illustrates one possible, significant pitfall in more
explicit semantics. As discussed above, such a semantics
might detail that a particular step in a computation causes the
spawning of a thread. 
More specifically,
it might extend transitions with an extra trace component: 
a triple
$(\sigma, u, \tau)$ might represent a step from $\sigma$ to $\tau$ that
spawns a thread that contains  the trace~$u$.
With such a semantics, the meanings of $\fork{(C
; \yield ; D)}$ and $\fork{(C; \fork{(D)})}$ would be different, since
they have different spawning behavior. 
\end{enumerate}

\noindent Many other useful equivalences hold. For instance,
we have:
\[
\dsem{\assign{x}{n} ; \assign{x}{n'}}
= 
\dsem{\assign{x}{n'}}
\]
trivially. For every $C$, we also have:
\[
\dsem{\fork{(C)}; \assign{x}{n}}
= 
\dsem{\assign{x}{n} ; \fork{(C)}}
\]
and, for every $C$ and $D$, we have:
\[
\dsem{\fork{(C)}; \fork{(D)}}
= 
\dsem{\fork{(D)}; \fork{(C)}}
\]
Another important equivalence is:
\[
\dsem{\Whiledo{(0=0)}{\unit}} = \dsem{\block}
\]
Thus, the semantics does not distinguish an infinite loop which never
yields from immediate blocking.  
On the other hand, we have:
\[
\dsem{\Whiledo{(0=0)}{\yield}} \neq \dsem{\block}
\]
The command $\Whiledo{(0=0)}{\yield}$ generates unbounded sequences 
of stutters $(\sigma,\sigma)$.
Similarly, we have:
\[
\dsem{{\yield}; {\yield}} \neq \dsem{\yield}
\]
Alternative 
semantics that would distinguish 
$\Whiledo{(0=0)}{\unit}$ from $\block$
or that would \hbox{identify}
$\Whiledo{(0=0)}{\yield}$ with $\block$
and
${\yield}; {\yield}$ with $\yield$
are viable, however.
We briefly discuss those variants and others in
Section~\ref{sec:conclusion}.

We leave as subjects for further research the problems of axiomatizing and of deciding equivalence and implementation relations, and the related  problem of  program verification, perhaps restricted to subsets of the language---even, for example, to the subset with just composition, spawning, and yielding. There is a large literature on axiomatization and decidability in concurrency theory; see, e.g.,~\cite{AI07} for discussion and further references.
Also, recent results on the automatic verification of asynchronous programs appear rather 
encouraging~\cite{majumdar07,majumdar09}; some of their ideas might be applicable in our setting.

\subsection{Two Extensions}

Trace-based semantics can also be given for variants and
enhancements of our basic imperative language. Here we illustrate this
point by considering two such enhancements, which illustrate the use
of $\finalm$ and $\donem$. Section~\ref{sec:conclusion} briefly
considers other possible language features.

\subsubsection{$\finishn$}

While cleaning maps a transition sequence
sequence to a proper pure transition sequence, a \emph{marking} 
function maps a proper pure transition sequence to a transition sequence.
For a proper pure transition sequence $u$, we define $\markret{u}$ by:
\[
\begin{array}{lclr}
\markret{v (\sigma, \sigma')\donem} & = & v(\sigma, \sigma' \ \finalm)\donem&\\
\markret{v} & = & v &\mbox{\rm \ (if $v$ does not contain $\donem$)}
\end{array}
\]
Thus, $\markret{u}$ includes a marker $\finalm$ only if $u$ contains a
marker $\donem$ (that is, if $u$ corresponds to a terminating
execution); the marker $\finalm$ is on the last transition of
$\markret{u}$, intuitively indicating that control is returned to the
sequential context when execution terminates.

Much as for cleaning, we extend marking to non-empty prefix-closed sets of proper pure transition sequences:
\[\markret{-}\type \PThread_{} \rightarrow \Proc_{}\] 
Using this extension, 
we can define the meaning of a construct $\finishn$, inspired by that of the X10 language~\cite{charles:x10,saraswat:x10}. 
We set: 
\[
\dsem{\finish{C}} = \markret{(\clean{\dsem{C}})}
\]

The intent is that $\finish{C}$ executes $C$ and returns control when all activities spawned by $C$ terminate. For instance, in $\finish{(\fork{(\assign{x}{0})})}; \assign{x}{1}$,
the assignment $\assign{x}{1}$ will execute only after $\assign{x}{0}$ is done.
In contrast, in 
$\fork{(\assign{x}{0})}; \assign{x}{1}$,
the assignments have the opposite ordering.
However, $\finish{(\fork{(\assign{x}{0})})}$
is not equivalent to
$\assign{x}{0}$, but rather to 
$\yield ; \assign{x}{0}$. Beyond
this simple example, $\finishn$ can be applied to more complex
commands, possibly with nested forks, and ensures that all the
activities forked terminate before returning control.

\subsubsection{Parallel Composition} \label{parcomp}

The definition of parallel composition relies on familiar themes: the
use of shuffling, and the decomposition of parallel composition into two cases.
The cases correspond to whether the left or the right argument of parallel composition takes the first step.

We define parallel composition at the level of transition sequences by
letting $u \parop u'$ and $u \parop_{l} u'$ be the least sets that satisfy prefix-closure
and the following clauses:
\begin{enumerate}[$\bullet$]
\item $w \in (\emptyseq \parop w)$ and $w \in  (w \parop \emptyseq)$,
\item $(t \parop_l t') \cup (t' \parop_l t) \subseteq (t \parop t')$,
\item if $v \in (w \parop t')$, then
$(\sigma,\sigma')v \in (\sigma,\sigma')w \parop_l t'$,
\item if $v \in \shuffle{w}{w'}$ then
$(\sigma,\tau)v \in (\sigma,\sigma' \ \finalm)w \parop_l (\sigma',\tau)w'$.
\end{enumerate}

\noindent Extending this function to 
\[{-}\parop{-}\type \Proc_{} \times \Proc_{} \rightarrow \Proc_{}\] 
we can define the meaning of a parallel-composition construct:
\[
\dsem{C\parop D} = \dsem{C} \parop \dsem{D}
\]
The reader may verify that parallel composition, as defined here, 
has the expected properties, for instance that it is commutative and associative with unit $\unit$.
It is also worth noting that  (under mild assumptions on the available expressions)
the binary nondeterministic choice operator $\cup$ considered in Section~\ref{Resumptions} is definable from parallel composition. The converse also holds, under restricted circumstances: if all occurrences of $\yield$ in $C$ and $D$ occur inside an $\fork\!\!$ then we have:
\[\dsem{C \parop D} = \dsem{C; D} \cup \dsem{D;C}\]
\section{Adequacy and Full Abstraction}
\label{sec:AdequacyFull}

In this section we establish that the denotational semantics of
Section~\ref{sec:dsemantics} coincides with the operational semantics
of Section~\ref{sec:osemantics}, and is fully abstract.

The adequacy theorem (Theorem~\ref{thm:adequacy}), which expresses the coincidence, 
says
that the traces that the denotational semantics predicts are exactly
those that can happen operationally.  These traces may in general
represent the behavior of a command in a context.  
As a special case, 
the adequacy theorem applies to runs, which are
essentially traces that the command can produce on its own, 
i.e., with an
empty context. This special case is spelled out in Corollary~\ref{thm:adequacy-runs} which states that the runs that the denotational semantics predicts are exactly
those that can happen operationally

The full-abstraction theorem (Theorem~\ref{thm:fullabs}) states that two commands $C$ and $D$
have the same set of traces denotationally if, and only if, they produce the same
runs in combination with every context. In particular, observing
runs, we cannot distinguish $C$ and $D$ in any context. Note that, given Corollary~\ref{thm:adequacy-runs}, we may equivalently speak of runs denotationally or operationally. We comment on other possible notions of
observation, and the corresponding full-abstraction results, below.

Section~\ref{sec:Runs} defines runs precisely. Sections~\ref{sec:Adequacy}
and~\ref{sec:fullabstractionII} present our adequacy and full-abstraction
results, respectively.

\subsection{Runs}
\label{sec:Runs}

A pure transition sequence  \emph{generates a run} 
if, however it can be written as
$u(\sigma,\sigma')(\sigma'',\sigma''') v$,
we have  $\sigma' = \sigma''$. If $w = (\sigma_1,\sigma_2)\dots (\sigma_{n-1},\sigma_{n})$ is such a pure transition sequence,  
we set $\onerun{w} = \sigma_1\ldots\sigma_{n}$ and $\onerun{w\,\donem} = \sigma_1\ldots\sigma_{n}\donem$.
A transition sequence $u$   \emph{generates a run} if $\clean{u}$ does,
and then
we set 
$\onerun{u} = \onerun{\clean{u}}$.

If a pure transition sequence $u$ generates a run, then 
it can be
easily be recovered from $\onerun{u}$: the run $\sigma_1 \ldots \sigma_{n}$
maps back to 
\[(\sigma_1, \sigma_2) \ldots (\sigma_{n-1}, \sigma_{n})
\] 
and the run
$\sigma_1 \ldots \sigma_{n} \donem$
maps back to 
\[
(\sigma_1, \sigma_2) \ldots (\sigma_{n-1}, \sigma_{n})\donem
\]
Since each non-empty run contains at least two elements, 
this definition applies
when $n = 0$ and $n \geq 2$.
We write $\runs{P}$ for the set of runs generated
by (pure) transition sequences in $P$.

\subsection{Adequacy} 
\label{sec:Adequacy}

\begin{lem} \label{equivalences} The following equalities hold:
\begin{enumerate}[\em(1)]
\item $\dsem{\avcxt{\block}} = \dsem{\block}$
\item $\dsem{\unit;C} = \dsem{C}$
\item $ \dsem{\avcxt{\fork{D}}} = \async(\clean{\dsem{D}},\dsem{\avcxt{\unit}}) $
\item $\clean{\dsem{\avcxt{\yield}}} = \clean{\async(\clean{\dsem{\avcxt{\unit}}},\dsem{\unit})} $
\item For all  $T \neq \emptythread$ (equivalently $\donem \not\in \dsem{T}$), 
\[\dsem{T} = \bigcup \{ \clean{\dsem{T'.T'', C}} \mid T = T'.C.T''\}\]
\end{enumerate}
\end{lem}
\proof
The first part is immediate from the semantics of $\block$ and the definition of composition. The second part holds as $\compunit$ is a unit for composition. The third part follows from the facts that $\async(P) \comp Q = \async(P,Q)$ and that composition is associative with unit $\compunit$.

For the fourth part, using the third part one sees that it is enough to show that for every $\avcxtsym$ we have:
\[ \clean{\dsem{\avcxt{\yield}}} = \clean{\dsem{\fork{\avcxt{\unit}}}}\]
As composition is associative with unit $\compunit$, this is equivalent to showing that, for every  $C$ we have:
\[\clean{\dsem{\yield; C}} = \clean{\async(\clean{\dsem{C}})}\]
which follows immediately, expanding the definitions. The proof of the fifth part is a straightforward verification.
\qed

\begin{lem} \label{immediately_blocked} If 
$C$ is blocked then,
for all $T$, $\dsem{T,C}=\{\emptyseq\}$.
\end{lem}

\proof We calculate:

    \[\begin{array}{lclr}
     	\dsem{T,\avcxt{\block}} & = & \async(\dsem{T},\dsem{\avcxt{\block}}) &\\
                              & = & \async(\dsem{T},\dsem{\block}) & \mbox{ (by Lemma~\ref{equivalences})}\\
                              & = & \{\emptyseq\} &
     \end{array}\]
\qed

\begin{lem} \label{skip} $\dsem{T,\unit}= \prefixclosure{\{(\sigma,\sigma \ \finalm)v\mid v \in  \dsem{T}\}}$.
\end{lem}
\proof Immediate from the definition of $\async$.
\qed

The next lemma applies when $C$ is neither $\unit$ nor blocked.
\begin{lem} \label{neither_skip_{}or_blocked}
	Suppose that $\nt{\sigma,T,C}\standardsaction\nt{\sigma',T',C'}$.
	Then, for any $\sigma''$,  $(\sigma,\sigma'')v \in \clean{\dsem{T,C}}$ iff $(\sigma', \sigma'')v \in \clean{\dsem{T',C'}}$.
\end{lem}
\proof We divide into cases according to the form of $C$. In the case where $C$ has the form $\avcxt{\unit;D}$ we have $\sigma' = \sigma$, $T' = T$ and $C' = \avcxt{D}$. So, by Lemma~\ref{equivalences}, we have $\dsem{T',C'} = \dsem{T,C}$, and we are done.
	
In the case where $C$ instead has the form $\avcxt{\fork{D}}$, we have $\sigma' = \sigma$, $T' = T.D$ and $C' = \avcxt{\unit}$ and we calculate:
		\[\begin{array}{lclr}
	     	{\dsem{T',C'}} & = & {\dsem{T.D,\avcxt{\unit}}}&\\
	                     & = & {\async(\dsem{T},\async(\clean{\dsem{D}},\dsem{\avcxt{\unit}}))} & \\
	                     & = & \dsem{T,\avcxt{\fork{D}}} & \mbox{ (by Lemma~\ref{equivalences})}\\
	                     & = & \dsem{T,C}
	     \end{array}\]
	and we are done.

In the case where $C$ instead has the form $\avcxt{\yield}$, we have $\sigma' = \sigma$, 
$T' = T.\avcxt{\unit}$, $C' = \unit$ and, again using  Lemma~\ref{equivalences}, we calculate:
		\[ \begin{array}{lclr}
		     \clean{\dsem{T',C'}} & = & \clean{\dsem{T.\avcxt{\unit},\unit}}&\\
	                     & = & \clean{\async(\dsem{T},\async(\clean{\dsem{\avcxt{\unit}}},\dsem{\unit})) }& \\
	                     & = & \clean{\dsem{T,\avcxt{\yield}}} &\\
	                     & = & \clean{\dsem{T,C}}
	         \end{array}
		\]
		and we are done.
		
In the next case, $C$ has the form $\avcxt{\assign{x}{e}}$, and we have $\sigma' = \sigma[x \mapsto \sigma(e)]$,  $T' = T$ and $C' = \avcxt{\unit}$. Here
 $\dsem{T,C} = \dsem{T, \assign{x}{e};\avcxt{\unit}}$. So we have that:
	$(\sigma,\tau)v \in \dsem{T,C}
	$ holds iff $(\sigma',\tau) \in \dsem{T,\avcxt{\unit}}$

Otherwise, $C$ has one of the forms $\avcxt{\Ifthenelse{b}{C}{D}}$ or 	
$\avcxt{\Whiledo{b}{C}}$ and we proceed much as in the previous case.
\qed

\begin{lem} \label{soundness} Suppose that $\nt{\sigma,T,C} \standardsaction^* \mbox{ some } \nt{\sigma',T',\unit}$ with $u \in \clean{\dsem{T'}}$. Then $(\sigma,\sigma')u \in \clean{\dsem{T,C}}$.
\end{lem}
\proof This follows from Lemmas~\ref{skip} and ~\ref{neither_skip_{}or_blocked}.
\qed

For the proof of the  converse of this lemma, we proceed by an induction on the size of loop-free commands.
We then extend to general commands by expressing their semantics in terms  of the semantics  of their approximations by loop-free commands. The \emph{size} of a loop-free command is defined by structural recursion: 
\[
\begin{array}{c}
\mymod{\unit} = \mymod{\block} = 1 \;\;\; \mymod{\assign{x}{e}} = \mymod{\fork{C}} = \mymod{\yield} = 2\\[.2em]
\mymod{\Ifthenelse{b}{C}{D}} = \mymod{C;D} = \mymod{C} + \mymod{D}
\end{array}
\] 
Note that if $\nt{\sigma,T, C} \standardsaction \nt{\sigma',T', C'}$ and $C$ is loop-free, then so is $C'$ and, further, $\mymod{C'} <  \mymod{C}$.

The \emph{approximation} relation $C \cmdapprox D$ between loop-free commands $C$ and general commands $D$ is defined to be the least such relation closed under all non-looping program constructs and such that, for any $b$, $C$, $D$, and $i \geq 0$:
\[\block \cmdapprox D\quad\quad\frac{C \cmdapprox D}{(\Whiledo{b}{C})_i \cmdapprox  (\Whiledo{b}{D})}\]
This relation is extended to thread pools and contexts in the obvious way: we write $T \cmdapprox T'$ and $\mathcal{C} \cmdapprox\mathcal{C}'$ for these extensions.

\begin{lem} \label{approximation1}
Suppose that $T \cmdapprox U$, $C \cmdapprox D$, and, further, that  $\nt{\sigma,T,C} \standardsaction \nt{\sigma',T',C'}$. Then, for some $U',D'$ with $T' \cmdapprox U'$ and $C' \cmdapprox D'$,  
 $\nt{\sigma,U,D} \standardsaction^* \linebreak[0] \nt{\sigma',U',D'}$.
 \end{lem}
\proof
One first notes that, for any $C$, $D$, if $\avcxt{C} \cmdapprox D$ then $D$ has the form  ${\mathcal E}'[D']$ where 
${\mathcal E} \cmdapprox {\mathcal E}'$ and $C \cmdapprox D'$. The proof then divides into cases according to the rule used to show that  $\nt{\sigma,T,C} \standardsaction \nt{\sigma',T',C'}$. 

For example, suppose we have 
$C = \avcxt{\Ifthenelse{b}{C_1}{C_2}}$ and  $\sigma(b)= \btrue $. We know that $D$ must have the form
${\mathcal E}'[D']$ where 
${\mathcal E} \cmdapprox {\mathcal E}'$ and $(\Ifthenelse{b}{C_1}{C_2}) \cmdapprox D'$. Suppose now that $D'$ has the form  $\Whiledo{b}{D''}$. Then we must have, for some $i \geq 0$ that 
$C_1 = C''; (\Whiledo{b}{C''})_i$ where $C'' \cmdapprox D''$. But then we observe that 
 \[\nt{\sigma,U,D} \standardsaction \nt{\sigma,U,{\mathcal E}'[\Ifthenelse{b}{D'';D'}{\unit}]}
 \standardsaction  \nt{\sigma,U,{\mathcal E}'[D'';D]}
 \]
 and the conclusion follows. The other cases  are straightforward.
\qed

Next we define the \emph{approximants} $C^{(i)}$ of a command $C$ by induction on $i$ and  structural 
recursion on $C$, beginning with the case where $C$ has one of the forms $\unit$, $\block$, $\assign{x}{e}$, or $\yield$, when $C^{(i)} = C$,  and continuing with:
\[
\begin{array}{r@{\ }c@{\ }l}
(\fork{C})^{(i)} &=& \fork{C^{(i)}}\\
(\Ifthenelse{b}{C}{D})^{(i)} &=& \Ifthenelse{b}{C^{(i)}}{D^{(i)}}\\
(C;D)^{(i)} &=& C^{(i)};D^{(i)}\\
(\Whiledo{b}{C})^{(i)} &=& (\Whiledo{b}{C^{(i)}})_{i}
\end{array}
\]
For any $C$ one shows that $C^{(i)} \cmdapprox C^{(i+1)} \cmdapprox C$.

\begin{lem} \label{approximation2}
$\mbox{}$
\begin{enumerate}[\em(1)]
       \item  If $C \cmdapprox D$ then $\dsem{C} \subseteq \dsem{D}$.
       \item  For any command $D$: 
\[\dsem{D} = \bigcup_i \dsem{D^{(i)}}\]
   \end{enumerate}
   \end{lem}
\proof The first part is evident using the monotonicity of the semantics of the program constructors and the semantic of loops. For the second part, we proceed by structural induction on $D$. All cases are straightforward, using the continuity of the program constructors, except for loops where we calculate:
	\[\begin{array}{lcl}
	\dsem{\Whiledo{b}{D}} & = & \bigcup_i \dsem{(\Whiledo{b}{D})_i}\\
	                      & = & \bigcup_i \dsem{(\Whiledo{b}{[\;]})_i}(\dsem{D})\\
		                  & = & \bigcup_i\dsem{(\Whiledo{b}{[\;]})_i}
		                           (\bigcup_i \dsem{D^{(i)}})\\
        		          & = & \bigcup_i \dsem{(\Whiledo{b}{D^{(i)}})_i}\\            
		                  & = &  \bigcup_i \dsem{(\Whiledo{b}{D})^{(i)}}
\rlap{\hbox to127 pt{\hfill\qEd}}
	\end{array}
	\]\unskip

We can now establish the converse of Lemma~\ref{soundness}.

\begin{lem} \label{progress} Suppose that $(\sigma,\sigma')u \in \clean{\dsem{T,C}}$. Then
$\nt{\sigma,T,C} \linebreak[0]\standardsaction^*  \nt{\sigma',T',\unit}$ for some $T'$ with $u \in \clean{\dsem{T'}}$.
\end{lem}
\proof We begin by proving this for loop-free commands $C$. The proof is by induction on the size of $C$. If $C$ is $\unit$ we have $\nt{\sigma,T,\unit} \standardsaction^* \nt{\sigma,T,\unit}$ and the conclusion follows, as, by Lemma~\ref{skip}, $(\sigma,\sigma')u \in \clean{\dsem{T,\unit}}$ iff $\sigma' = \sigma$ and $u \in \clean{\dsem{T}}$. If $C$ is blocked, the conclusion holds trivially, by Lemma~\ref{immediately_blocked}.
	
	If $C$ is neither $\unit$ nor blocked we have $\nt{\sigma,T,C} \standardsaction \nt{\sigma'',T'',C''}$ (and then $C''$ is loop-free and $\mymod{C''} < \mymod{C}$). 
	Then, by Lemma~\ref{neither_skip_{}or_blocked}, $(\sigma,\sigma')u \in \clean{\dsem{T,C}}$ 
iff $(\sigma'',\sigma')u \in \clean{\dsem{T'',C''}}$ which latter, by the induction hypothesis, implies $\nt{\sigma'',T'',C''} \standardsaction^* \mbox{ some } \nt{\sigma',T',\unit}$ with $u \in \clean{\dsem{T'}}$ which, in turn, implies $\nt{\sigma,T,C} \standardsaction^* \mbox{ some } \nt{\sigma',T',\unit}$ with $u \in \clean{\dsem{T'}}$, as desired.

Next suppose that $(\sigma,\sigma')u \in \clean{\dsem{T,D}}$, where now $D$ 
is not  loop-free. By Lemma~\ref{approximation2} 
$(\sigma,\sigma')u \in \clean{\dsem{T,C}}$ for some $C \cmdapprox D$.
So, by the above,  $\nt{\sigma,T,C} \standardsaction^* \mbox{ some } \nt{\sigma',T',\unit}$ with $u \in \clean{\dsem{T'}}$. The desired conclusion follows immediately, using Lemma~\ref{approximation1}.
\qed

\begin{lem} \label{realprogress} 
$\mbox{}$
\begin{enumerate}[\em(1)]
\item For any proper non-empty pure transition sequence $u$,  $(\sigma,\sigma')u \in \clean{\dsem{T,C}}$ iff for some $T',C'$,
$\nt{\sigma,T,C} \linebreak[0]\standardsaction^*\standardschoice  \nt{\sigma',T',C'}$ with $u \in \clean{\dsem{T',C'}}$.
\item For any $\sigma$, $\sigma'$, $T$, $C$, $(\sigma,\sigma')\donem \in \clean{\dsem{T,C}}$ iff 
$\nt{\sigma,T,C}\linebreak[0]\standardsaction^* \nt{\sigma',\emptythread,\unit}$.
\end{enumerate} 
\end{lem}
\proof
By Lemma~\ref{progress}, $(\sigma,\sigma')u \in \clean{\dsem{T,C}}$ holds iff $\nt{\sigma,T,C} \standardsaction^* \mbox{  some  } \nt{\sigma',T',\unit}$ does, with $u \in \dsem{T'}$. In the case where $u$ is proper the conclusion follows from  Lemma~\ref{equivalences}. In the case where $u$ is $\donem$ we see from the definition of $\dsem{T'}$ 
that	$\donem \in \dsem{T'}$ iff $T' = \emptythread$.
\qed

The following \emph{Adequacy Theorem} for pure transition sequences is an immediate consequence of Lemmas~\ref{progress} 
and~\ref{realprogress}:
\begin{thm}\label{thm:adequacy}
$\mbox{}$
\begin{enumerate}[\em(1)]
	\item For $n > 0$, $(\sigma_1,\sigma_1')\ldots (\sigma_{n},\sigma_{n}') \in \clean{\dsem{T,C}}$ iff there are $T_i,C_i$, ($i = 1,n$) such that $T_1 = T$, $C_1 = C$, and  $\nt{\sigma_i,T_i,C_i} \standardsaction^*\standardschoice \nt{\sigma_{i}',T_{i+1},C_{i+1}}$, for $1 \leq i \leq n-1$, and $\nt{\sigma_{n},T_{n},C_{n}} \standardsaction^* \mbox{some  } \nt{\sigma_{n}',T', \unit}$.
	
	\item For $n > 0$,  $(\sigma_1,\sigma_1')\ldots (\sigma_{n},\sigma_{n}')\donem \in \clean{\dsem{T,C}}$ iff there are $T_i,C_i$, ($i = 1,n$) such that $T_1 = T$, $C_1 = C$, and $\nt{\sigma_i,T_i,C_i} \standardsaction^*\standardschoice \nt{\sigma_{i}',T_{i+1},C_{i+1}}$, for $1 \leq i \leq n-1$, and $\nt{\sigma_{n},T_{n},C_{n}} \standardsaction^* \nt{\sigma_{n}',\emptythread, \unit}$.\qed
\end{enumerate}	
\end{thm}

As a corollary we obtain an adequacy theorem for runs:
\begin{cor} \label{thm:adequacy-runs}
$\mbox{}$
\begin{enumerate}[\em(1)]
	\item  For $n \geq 2$, $\sigma_1\ldots\sigma_{n} \in \runs{\dsem{T,C}}$  iff there are $T_i,C_i$, ($i = 1,n-1$) such that $T_1 = T$, $C_1 = C$, $\nt{\sigma_i,T_i,C_i} \standardsaction^*\standardschoice \nt{\sigma_{i+1},T_{i+1},C_{i+1}}$ ($ 1 \leq i \leq n-2$), and $\nt{\sigma_{n-1},T_{n-1},C_{n-1}} \standardsaction^* \mbox{some  } \nt{\sigma_{n},T', \unit}$.
	
	\item For $n\geq 2$, $\sigma_1\ldots \sigma_{n}\donem  \in \runs{\dsem{T,C}}$  iff there are $T_i,C_i$, ($i = 1,n-1$) such that $T_1 = T$, $C_1 = C$, and $\nt{\sigma_i,T_i,C_i} \linebreak[0]\standardsaction^*\standardschoice \nt{\sigma_{i+1}T_{i+1},C_{i+1}}$ ($ 1 \leq i \leq n-2 $), and $\nt{\sigma_{n-1},T_{n-1},C_{n-1}} \linebreak[0]\standardsaction^* \nt{\sigma_{n},\emptythread, \unit}$.\qed
\end{enumerate}	
\end{cor}


\subsection{Full Abstraction}
\label{sec:fullabstractionII}

The first lemma in the proof of full abstraction bounds the
nondeterminism of commands in semantic terms.  

\begin{lem}\label{lemma:fin} 
For all $C$, $u$, and $\sigma$, the set $\{ \tau \mid u(\sigma, \tau) \in \dsem{C}\}$ is finite.
\end{lem}
\proof
More generally, we prove that 
for all $T$, $C$, $u = (\sigma_1,\tau_1)\ldots(\sigma_{n-1},\tau_{n-1})$, and $\sigma_{n}$, the set $\{ \tau_{} \mid u(\sigma_{n}, \tau_{}) \in \dsem{T,C}\}$ is finite, and similarly that 
the set $\{ \tau_{} \mid u(\sigma_{n}, \tau_{}) \in \dsem{T}\}$ is finite.
The proof is by induction on $n$.
The proof relies on adequacy; a purely semantic proof might be possible but seems harder.
\begin{enumerate}[$\bullet$]

\item If $C$ is $\unit$, then Lemma~\ref{skip} implies that 
$\tau_1$ is $\sigma_1  \ \finalm$, and  
$(\sigma_2,\tau_2)\ldots(\sigma_{n},\tau_{}) \in \dsem{T_1}$.
In case $n=1$, we are done, with a unique choice for $\tau_1$.
Otherwise, we conclude by induction hypothesis.

\item if $C$ is blocked, then $n=0$, by Lemma~\ref{immediately_blocked}, 
so this case is vacuous.

\item If $C$ is neither $\unit$ nor blocked, then Lemma~\ref{progress} implies that $\tau_1$ is unique.
In case $n=1$, we are done, with a unique choice for $\tau_1$.
Otherwise, Lemma~\ref{progress} also implies that 
$(\sigma_2,\tau_2)\ldots(\sigma_{n},\tau_{}) \in \dsem{T'}$ for a unique $T'$.
As in the case of $\unit$, the desired conclusion follows by induction hypothesis.

\item Finally, having established the claim for sequences
of length $n$ for sets of the form $\dsem{T,C}$, 
we consider sequences of length $n$ in a set of the form $\dsem{T}$.
Suppose that $T$ consists of $C_1, \ldots, C_k$.
A transition sequence $v$ in $\dsem{T}$ is a shuffle of transition sequences
in $\dsem{C_1}$,\ldots,$\dsem{C_k}$, each of length at most $n$. 
The finiteness property for 
$\dsem{T}$ follows from the fact that there are only finitely many possible ways of decomposing $v$ as a shuffle.\qed
\end{enumerate}

Intuitively, Lemma~\ref{lemma:fin} 
is useful because it implies that, at any point, there are certain
steps that a command cannot take, and in proofs those steps can be used as
unambiguous, visible markers of activity by the context.
This lemma is somewhat fragile---it does not hold once 
one adds to the language either the nondeterministic choice operator considered in Section~\ref{Resumptions}  or the parallel composition operator of Section~\ref{parcomp}. It follows that neither of these operators is definable in the language.
An alternative argument that does
not use the lemma relies on fresh variables instead. The fresh
variables permit an alternative definition of the desired markers.

Full-abstraction results invariably require some notion of
observation.  Let us write $\observ{P}$ for the
observations that we make on $P \in \Proc_{}$.  Equational full abstraction is
that $\dsem{C} = \dsem{D}$ if and only if, for every context
$\ccxtsym$, 
we have  
$\observ{\dsem{\ccxt{C}}} = \observ{\dsem{\ccxt{D}}}$.  
In other words, 
two commands have the same meaning if and only if they
yield
the same observations in every context of the language.  
The stronger inequational full abstraction is that $\dsem{C} \subseteq \dsem{D}$
if and only if, for every context $\ccxtsym$, 
we have
$\observ{\dsem{\ccxt{C}}} \subseteq \observ{\dsem{\ccxt{D}}}$.  The
difficult part of this equivalence is usually the implication from
right to left: that if, for every context $\ccxtsym$,
$\observ{\dsem{\ccxt{C}}} \subseteq \observ{\dsem{\ccxt{D}}}$, then
$\dsem{C} \subseteq \dsem{D}$.

One possible candidate for $\observ{P}$ is $\clean{P}$. This notion of
observation can be criticized as too fine-grained. Nevertheless, we
find it useful to prove full abstraction for this notion of observation,
with the following lemma.
We first need some auxiliary definitions for its proof, and the lemma that follows.
Given two stores $\sigma$ and $\sigma'$, we define:
\begin{enumerate}[$\bullet$]
\item  a boolean expression $\checkstore{\sigma}$
as the conjunction of the formulas $x = n$ for every 
variable $x$, where $n$ is the natural number $\sigma(x)$ (so
$\checkstore{\sigma}$ is true in $\sigma$ and false elsewhere);

\item a command $\gotostore{\sigma}$ as the sequence of assignments $\assign{x}{n}$ for every 
variable $x$, where $n$ is the natural number $\sigma(x)$;

\item a command
$\gofromtostore{\sigma}{\sigma'}$ as
$\Ifthenelse{\checkstore{\sigma}}{\gotostore{\sigma'}}{\linebreak[0]\block}$;

\item a command
$\tgofromtostore{\sigma}{\sigma'}{\sigma''}$ as
$\gofromtostore{\sigma}{\sigma'}; \yield ;
\gofromtostore{\sigma'}{\sigma''}; \yield$.

\end{enumerate}%
These definitions exploit the fact that the set of variables is
finite. However, with more care, analogous definitions could be given otherwise, by focusing on the set of variables relevant to the programs under observation.

\begin{lem}\label{lemma:clean}
If 
$\clean{\dsem{\ccxt{C}}}  \subseteq
\clean{\dsem{\ccxt{D}}}$ for every context $\ccxtsym$, then
$\dsem{C} \subseteq\dsem{D}$.
\end{lem}
\proof
Letting $P = \dsem{C}$ and $Q = \dsem{D}$, we assume that $P \not\subseteq Q$ and
prove that there exists $\ccxtsym$ such that 
$\clean{\dsem{\ccxtsym}(P)} \not\subseteq
\clean{\dsem{\ccxtsym}(Q)}$.
For this, choose a sequence $w$ 
in $P$ but not in $Q$.
If $w = \clean{w}$, then we can take $\ccxtsym$ to be $\hole$.
Therefore, for the rest of the proof, we consider the case
$w \neq \clean{w}$.

If $w \neq \clean{w}$, then $w$ is of the form 
$u (\sigma, \sigma'\  \finalm) v$.
We let $\ccxtsym = \hole ; \gofromtostore{\sigma'}{\sigma''}$ where
$\sigma''$ does not appear in $u$ or $v$ and 
$u (\sigma, \sigma'') \not\in Q$
(so, by prefix-closure, 
$u (\sigma, \sigma'')v \not\in Q$). 
Such a choice of $\sigma''$ is always
possible by Lemma~\ref{lemma:fin}.
Thus, 
$\dsem{\ccxtsym}(P)$ contains $u (\sigma, \sigma''\  \finalm) v$,
and 
$\clean{\dsem{\ccxtsym}(P)}$ contains $u (\sigma, \sigma'') v$.

Suppose that $u (\sigma, \sigma'') v$
is also in 
$\clean{\dsem{\ccxtsym}(Q)}$,
and that this is because some sequence $w'$
is in $\dsem{\ccxtsym}(Q)$ and
$\clean{w'} = u (\sigma, \sigma'') v$.
By the definition of the semantics of sequential composition, 
this could arise in one of the following ways:
\begin{enumerate}[$\bullet$]

\item $w' = u (\sigma, \sigma''\  \finalm) v$, with 
$w \in Q$. This contradicts $w \not\in Q$.

\item $w' = u' (\sigma, \sigma'') v'$,
and $\sigma''$ occurs as the second store of a {\final} transition in either $u'$
or $v'$. 
This contradicts the requirement that $\sigma''$ does not appear in $u$ or $v$.

\comment{GDP I missed out   ``where $\clean{u'} = u$ and $\clean{v'} = v$''. If that is not liked, one  could write $\clean{(u'v')} = uv$ instead}

\item $w' = u (\sigma, \sigma'') v$,
$w' \in Q$,  and $w'$ does not have a {\final}   transition.
This contradicts the requirement
that $u (\sigma, \sigma'') \not\in Q$.\qed
\end{enumerate}

Another possible candidate for $\observ{P}$ is $\runs{P}$.  Runs
record more than mere input-output behavior, but much less than 
entire execution histories. We therefore find them attractive for our purposes.
The following lemma connects runs to cleaning.

\begin{lem}\label{lemma:runs}
If 
$\runs{\dsem{\ccxt{C}}} \subseteq
\runs{\dsem{\ccxt{D}}}$ for every context~$\ccxtsym$, 
then 
$\clean{\dsem{C}} \subseteq \clean{\dsem{D}}$.
\end{lem}
\proof
Letting $P = \dsem{C}$ and $Q = \dsem{D}$, 
we assume that $\clean{P} \not\subseteq \clean{Q}$ and
prove that there exists $\ccxtsym$ such that 
$\runs{\dsem{\ccxtsym}(P)} \not\subseteq
\runs{\dsem{\ccxtsym}(Q)}$.

For this, choose a sequence $w \in \clean{P}$ but $w \not\in \clean{Q}$, in order to derive a contradiction.

First, suppose that $w$ is of the form
$(\sigma_1,\sigma'_1)\ldots(\sigma_{n},\sigma'_{n})$, with $n > 0$.
We let $\ccxtsym$ be $\fork{\hole}; \mesh{w}$, where
$\mesh{w}$ is
the command
\[\begin{array}{l}
\yield ; 
\tgofromtostore{\sigma'_1}{\sigma''_1}{\sigma_2};
\ldots ;
\tgofromtostore{\sigma'_{n-1}}{\sigma''_{n-1}}{\sigma_{n}};
\gofromtostore{\sigma'_{n}}{\sigma''_{n}}
\end{array}
\]
where the stores
$\sigma''_i$ are all different from one another and from all other
stores in $w$, and are such that
\[(\sigma_1,\sigma'_1)\ldots(\sigma_i,\sigma'_i)(\sigma'_i,\sigma''_i) \not\in \clean{Q}
\]
and
\[(\sigma_1,\sigma'_1)\ldots
(\sigma_{i-1},\sigma'_{i-1})
(\sigma''_{i-1},\sigma_i)(\sigma_i,\sigma'_i)(\sigma'_i,\sigma''_i) \not\in \clean{Q}
\]
Such a choice of stores $\sigma''_i$ is always
possible by Lemma~\ref{lemma:fin}.
Since 
$\dsem{\mesh{w}}$ contains
the transition sequence:
\[
(\sigma_1,\sigma_1)
(\sigma'_1,\sigma''_1)
(\sigma''_1,\sigma_2)
\ldots
(\sigma''_{n-1},\sigma_{n})
(\sigma'_{n},\sigma''_{n}\  \finalm)
\donem
\]
we obtain that $\dsem{\ccxtsym}(P)$ contains 
the transition sequence:
\[
\begin{array}{l}
(\sigma_1,\sigma_1)
(\sigma_1,\sigma'_1)(\sigma'_1,\sigma''_1)
(\sigma''_1,\sigma_2)
(\sigma_2,\sigma'_2)\ldots
(\sigma''_{n-1},\sigma_{n})
(\sigma_{n},\sigma'_{n})
(\sigma'_{n},\sigma''_{n}\  \finalm)
\end{array}
\]
which generates the run:
\[
\sigma_1
\sigma_1 \sigma'_1 \sigma''_1
\sigma_2
\sigma'_2\ldots
\sigma''_{n-1}
\sigma_{n}
\sigma'_{n}
\sigma''_{n}
\]
Suppose that this run is also in $\runs{\dsem{\ccxtsym}(Q)}$. 
Therefore, there exists $w' \in \clean{Q}$ such that 
\[
(\sigma_1,\sigma'_1)(\sigma'_1,\sigma''_1)
(\sigma''_1,\sigma_2)
(\sigma_2,\sigma'_2)\ldots
(\sigma''_{n-1},\sigma_{n})
(\sigma_{n},\sigma'_{n})
(\sigma'_{n},\sigma''_{n})
\]
is a shuffle of $w'$ with
\[
(\sigma'_1,\sigma''_1)
(\sigma''_1,\sigma_2)
\ldots
(\sigma''_{n-1},\sigma_{n})
(\sigma'_{n},\sigma''_{n})
\donem
\]
which we call $w''$, 
or with a prefix of $w''$.  
We analyze the origin of the transitions in the shuffle:
\begin{enumerate}[$\bullet$]
\item The transitions
$(\sigma_i,\sigma'_i)$ must all come from $w'$, since each of the
transitions in $w''$ contains one of the stores
$\sigma''_j$ and, by choice, these are different from $\sigma_i$ and
$\sigma'_i$.

\item Suppose that, up to some $i-1 < n$, $w'$ starts like $w$, in
other words 
it starts as
$(\sigma_1,\sigma'_1)\ldots(\sigma_{i-1},\sigma'_{i-1})$.  Suppose
further that, in the shuffle up to this point, each transition
$(\sigma_j,\sigma'_j)$ is followed immediately by the corresponding
transitions $(\sigma'_j,\sigma''_j)(\sigma''_j,\sigma_{j+1})$ from $w''$.
We argue that this remains the case up to $n$.

\begin{enumerate}[$-$]
\item We consider  $(\sigma'_{i-1}, \sigma''_{i-1})$, the next possible transition in the shuffle. This transition cannot come from $w'$ because, by the choice of $\sigma''_{i-1}$, we have that
\[(\sigma_1,\sigma'_1)\ldots(\sigma_{i-1},\sigma'_{i-1})(\sigma'_{i-1}, \sigma''_{i-1}) \not\in \clean{Q}
\] 
So this transition comes from $w''$.

\item One step further, in order to 
derive a contradiction,
we suppose that the transition $(\sigma''_{i-1},
\sigma_i)$ comes from $w'$. So $w'$ starts:
\[(\sigma_1,\sigma'_1)\ldots(\sigma_{i-1},\sigma'_{i-1})(\sigma''_{i-1}, \sigma_i)\] 
and in fact:
\[(\sigma_1,\sigma'_1)\ldots(\sigma_{i-1},\sigma'_{i-1})(\sigma''_{i-1}, \sigma_i)(\sigma_i,\sigma'_i)\]
 since, as noted above, the last
transition here must come from $w'$.
The next transition in the shuffle is $(\sigma'_i, \sigma''_i)$.
By the choice of $\sigma''_i$, we have that 
\[(\sigma_1,\sigma'_1)\ldots
(\sigma_{i-1},\sigma'_{i-1})
(\sigma''_{i-1},\sigma_i)(\sigma_i,\sigma'_i)(\sigma'_i,\sigma''_i) \not\in \clean{Q}
\]
So the transition $(\sigma'_i, \sigma''_i)$ cannot come from
$w'$. Therefore, it must come from $w''$. However, the next available
transition in $w''$ is $(\sigma''_{i-1}, \sigma_i)$, and $(\sigma'_i,
\sigma''_i)$ and $(\sigma''_{i-1}, \sigma_i)$ must be different
because $\sigma''_{i-1}$ and $\sigma''_i$ are different, by choice,
from $\sigma'_i$ and $\sigma_i$.

Thus, the assumption that the
transition $(\sigma''_{i-1}, \sigma_i)$ comes from $w'$ leads to a
contradiction. This transition must come from $w''$.
\end{enumerate}

\item Finally, suppose that, up to $n$, $w'$ starts like $w$, in
other words as:
\[(\sigma_1,\sigma'_1)\ldots(\sigma_{n},\sigma'_{n})\]
and
that, in the shuffle, each transition $(\sigma_j,\sigma'_j)$ is
followed immediately by the corresponding transitions
$(\sigma'_j,\sigma''_j)(\sigma''_j,\sigma_{j+1})$ from $w''$.  
By the choice of $\sigma''_{n}$, we have that 
\[(\sigma_1,\sigma'_1)\ldots(\sigma_{n},\sigma'_{n})(\sigma'_{n},\sigma''_{n}) \not\in \clean{Q}
\]
so $(\sigma'_{n},\sigma''_{n})$ comes from $w''$, not from $w'$.
\end{enumerate}
In sum, $w' = w$, and therefore $w \in \clean{Q}$, contradicting our assumption that $w \not\in \clean{Q}$.

Next, suppose that $w$ is of the form
$(\sigma_1,\sigma'_1)\ldots(\sigma_{n},\sigma'_{n})\ \donem$.
With the same
$\ccxtsym$, we obtain that 
$\dsem{\ccxtsym}(P)$ contains the transition sequence:
\[
\begin{array}{l}
(\sigma_1,\sigma_1)
(\sigma_1,\sigma'_1)(\sigma'_1,\sigma''_1)
(\sigma''_1,\sigma_2)
(\sigma_2,\sigma'_2)\ldots
(\sigma''_{n-1},\sigma_{n})
(\sigma_{n},\sigma'_{n})
(\sigma'_{n},\sigma''_{n}\  \finalm)
\donem
\end{array}
\]
which generates the run:
\[
\sigma_1
\sigma_1 \sigma'_1 \sigma''_1
\sigma_2
\sigma'_2\ldots
\sigma''_{n-1}
\sigma_{n}
\sigma'_{n}
\sigma''_{n}
\donem
\]
Suppose that this run is also in $\runs{\dsem{\ccxtsym}(Q)}$. 
Again, 
by the choice of $\sigma''_1$, \ldots, $\sigma''_{n}$,  
this can be the case only if 
$w$ is in $\clean{Q}$. (The argument for the contradiction may
actually be simplified in this case, because of the marker $\donem$.)
\qed

We obtain the following \emph{Full-abstraction Theorem}:

\begin{thm}\label{thm:fullabs}
$\dsem{C} \subseteq \dsem{D}$ iff, for every context $\ccxtsym$,
$\runs{\dsem{\ccxt{C}}} \linebreak[0]\subseteq  
\runs{\dsem{\ccxt{D}}}$.
\end{thm}
\proof

The implication from $\dsem{C} \subseteq \dsem{D}$ is an immediate consequence of the
compositionality of the semantics (Proposition~\ref{prop:comp}). The converse  follows from Lemmas~\ref{lemma:clean}
and~\ref{lemma:runs}. \qed

Coarser-grained definitions of $\observ{P}$
may sometimes be appropriate. 
For those, we expect that 
full abstraction will typically require additional
closure conditions on $P$, such
as closure under 
suitable forms of
stuttering and 
mumbling,
much as in our work and
Brookes's on parallel
composition~\cite{TCS::AbadiP1993,IC::Brookes1996}.

\section{Algebra} 
\label{sec:algebra}

The development of the denotational semantics in Section~\ref{sec:dsemantics} is ad hoc, in that the semantics is not related to any systematic approach. In this section we show how it fits in with the algebraic theory of 
effects~\cite{PP1,PP2,HPP,DBLP:conf/lics/PlotkinP08,PP09}.

In the {functional programming} approach to imperative languages, commands have unit type, $1$. Then, taking the {monadic} point of view~\cite{BHM}, they are modeled as elements of $T(1)$ for a suitable monad $T$ on, say, the category of $\omega$-cpos and continuous functions. For parallelism one might look for something along the lines of the resumptions monad~\cite{HennessyM:fulaspl,CM,HPP}. 

In the {algebraic} approach to computational effects~\cite{PP1,HPP}, one analyses the monads as free algebra monads 
for a suitable equational or Lawvere theory $L$ (here 
meaning in the enriched sense, so that inequations are allowed, as are families of operations continuously parameterized over an $\omega$-cpo). The operations of the theory (for example a binary choice operation in the case of nondeterminism)  are thought of as effect constructors in that they create the effects at hand.

As discussed in~\cite{HennessyM:fulaspl}, resumptions are generally not fully abstract when their domain equation is solved in a category of cpos. If, instead, it is solved in a category of semilattices, increased abstraction may be obtained. The situation was analyzed from the algebraic point of view in~\cite{HPP}. It was shown there that resumptions arise by combining a theory for stores~\cite{PP1} with one for nondeterminism, one for nontermination, and one for a unary operation 
$\delay$ thought of as suspending computation.  
The difference between solving the 
equation in a category of semilattices or cpos essentially amounts to whether or not one asks that $\delay$, and the other operations,  commute with nondeterminism. 

In~\cite{IC::Brookes1996}, Brookes, using an apparently different and mathematically elementary trace-based approach, succeeded in giving a fully abstract semantics for a language of the kind considered in~\cite{HennessyM:fulaspl}. However, in~\cite{LICS::Jeffrey1995}, Jeffrey showed that trace-based models of concurrent languages can arise as solutions to domain equations in a category of semilattices, thereby relating the two approaches.

We propose here to identify the suspension operation $\delay$ with the operation of the same name introduced in Section~\ref{sec:semcommand}; indeed this identification was the origin of the definition of $\yield$ given there, and it is natural to further identify $\yield$ as the generic effect~\cite{PP2} corresponding to the suspension operation. These identifications 
are justified by Corollary~\ref{Proc-algebra}, 
below, and the discussion following~it. 

In Section~\ref{Resumptions} we carry out an algebraic analysis of resumptions. 
We show in Theorem~\ref{resumptions} that, imposing the commutations with nondeterminism just discussed,  they do indeed correspond to a traces model, provided one uses the Hoare or lower powerdomain. (This powerdomain is a natural choice as we consider only ``may'' semantics in this paper, and elements of such powerdomains are Scott closed, so downwards-closed, a natural generalization of our prefix-closedness condition.) The proof makes the link between domain equations and traces.

The missing ingredient in an algebraic analysis of $\Proc_{}$ is then an account of 
$\forkn$. 
In the denotational semantics of any command of the form  $\fork{C}$, all $ \finalm$ marking is lost from the meaning of $C$,  because of the application of the cleaning function, $-^c$; further all the  sequences in $\dsem{C}^c$ are proper. We propose to treat $\forkn$ as a generic effect, parameterized by an element of $\PThread_{}$
(which will be $\dsem{C}^c$).

In order to give the equations for the $\async$ operation it will, as one may  expect, be useful to first have an algebraic analysis of $\PThread$; we carry out  this analysis in Section~\ref{Threads}. It turns out, as detailed in Theorem~\ref{threads}, that $\PThread$ is similar to, but not quite, a resumptions $\omega$-cpo.
Finally,  we analyze processes in Section~\ref{Processes}, showing, in Theorem~\ref{processes}, that a process is a kind of ``double-thread''---more precisely, a resumption that returns not only a value but also an element of $\PThread$.

\subsection{Resumptions} \label{Resumptions}

Our theory $L_{\Res}$  for resumptions follows~\cite{HPP} but is somewhat modified, as we are interested only in ``may'' semantics and as we wish to allow infinitely proceeding processes. The theory is a combination of several constituent theories which we now
consider successively.

The Lawvere theory $L_{\mathrm{S}}$ of stores can be presented via a family of unary operations $\update_{x,n}$ and a family of  ``$\mathbb{N}$-ary'' operations $\lookup_{x}$ ($x \in \fun{Vars}$, $n \in \mathbb{N}$). (An $\mathbb{N}$-ary operation is a countably infinitary operation whose arguments are indexed by 
the natural numbers.) For any computation $\gamma$, $\update_{x,n}(\gamma)$ is read as the computation that first updates $x$ to $n$ and then proceeds as $\gamma$; for any $\mathbb{N}$-indexed collection  $(\gamma_{n})_{n}$ of computations, $\lookup_{x}((\gamma_{n})_{n})$ 
is read as the computation that proceeds as $\gamma_{n}$ if  $x$ has value $n$
in the current store.

The Lawvere theory $L_{\mathrm{H}}$
 for nondeterminism is that of the lower (aka Hoare) powerdomain, presented using a binary nondeterministic choice operation
 $\cup$; the Lawvere theory  $L_{\Omega}$ for nontermination is the theory  of a least element, presented using a constant $\Omega$; and the Lawvere theory $L_{\delay}$ for suspension  is that  of a unary operation $\delay$, with no equations. See ~\cite{PP1,HPP} for more details of these theories, including an account of the equations for stores and for Hoare powerdomains.

For resumptions, continuing to follow~\cite{HPP}, we wish the operations of $L_{\mathrm{S}}$ to commute with those of $L_{\mathrm{H}}$ and $L_{\Omega}$ (which automatically commute with each other) and it is also natural to have $\delay$ commute with nondeterministic choice, but not with the operations of $L_{\mathrm{S}}$, as we wish to model interruption points, and not with $\Omega$, as we want to be able to model infinitely proceeding processes. We therefore define:
\[L_{\Res} = L_{\mathrm{H}} \otimes ((L_{\mathrm{S}} \otimes L_{\Omega}) + L_{\delay})\]
and let $T_{\Res}$ be the associated monad.  (For any two theories $L$ and $L'$ presented using disjoint signatures, the theories $L + L'$ and $L \otimes L'$ can be presented using the union of the signatures of $L$ and $L'$ and, in the former case, by the union of their equations and, in the latter case, by the union of their equations together with additional equations that say that each operation of each theory commutes with each operation of the other.)

We now give an elementary trace-based picture of $T_{\Res}(P)$ for 
sufficiently
general 
$\omega$-cpos $P$. Let $Q$ be a partial order. A \emph{$Q$-transition} is a pair of states $(\sigma, \sigma'\ x)$ in which the second is marked with an element $x$ of $Q$;  we let $\tau$ range over stores and stores marked with an element of $Q$. A \emph{basic $Q$-transition sequence} is a non-empty sequence consisting of plain transitions optionally followed by a  $Q$-transition. 
Let $\leq_Q$ be the least preorder on the set  of basic $Q$-transition sequences which contains the prefix relation $\prefix$ and is such that, for any $x$, $y$ in $Q$, if $x \leq y$ then $u(\sigma, \sigma' x) \leq_Q u(\sigma, \sigma' y)$. One 
has that $\leq_Q$ is a partial order and that $u \leq_Q v$ holds iff:
\[\begin{array}{ll}
\mbox{either  } &  u \prefix v\\
\mbox{or else } &  \exists w, x \leq y.\, u \prefix w(\sigma, \sigma' x) \land v =  w(\sigma, \sigma' y)
\end{array}\]

We need a few notions concerning ideals in partial orders.
An \emph{ideal} $I$ in a partial order $Q$ is a downwards-closed subset of $Q$; 
for any subset $X$ of $Q$ we write $X\!\!\downarrow$ 
for the least ideal including $X$, viz $\{ x \in Q \mid \exists y \in X. \,x \leq y \}$;  and for any $x \in Q$ we write $x\!\!\downarrow$ for 
$\{x\}\!\!\downarrow$.   Downwards-closed sets, i.e., ideals, provide a suitable generalization of  prefix-closed sets when passing from sequences to general partial orders.

An ideal $I$
is \emph{directed} if it is nonempty and any two elements of the 
ideal have an upper bound in the ideal.
An ideal  is \emph{denumerably generated} if 
$I = X\!\!\downarrow$ for some denumerable $X \subseteq I$.
We write ${\Id}^{\uparrow}_{\omega}(Q)$, respectively ${\Id}_{\omega}(Q)$, 
for the collection of all 
denumerably generated directed ideals  of $Q$, 
respectively all  denumerably generated 
ideals  of $Q$, and 
we partially order them by subset; 
$\Id^{\uparrow}_{\omega}(Q)$ is an $\omega$-cpo, indeed it is the free such over $Q$;
and ${\Id}^{\omega}(Q)$ is the free  $\omega$-cpo with all finite sups over $Q$: 
it follows that it is also the free such $\omega$-cpo over $\Id^{\uparrow}_{\omega}(Q)$.

Let $Q$-$\BTrans$ be the set of basic $Q$-transition sequences, partially ordered as above. One can view  $\Id^{}_{\omega}(\mbox{$Q$-$\BTrans$})$ as an $L_{\Res}$-model with the following definitions of the operations, where now we use $l$ to range over $\fun{Vars}$:

\[
\begin{array}{rcl}
(\update_{l,n})_{\Id^{}_{\omega}(\mbox{$Q$-$\BTrans$})}(I) &=& \{(\sigma,\tau)u \mid (\sigma[l \mapsto n],\tau)u \in I \}\\
(\lookup_l)_{\Id^{}_{\omega}(\mbox{$Q$-$\BTrans$})} ((I_{n})_{n}) &=& \bigcup_n \{(\sigma,\tau)u \in I_{n} \mid \sigma(l) = n\}  \\
I \cup_{\Id^{}_{\omega}(\mbox{$Q$-$\BTrans$})} J &=& I \cup J\\
\Omega_{\Id^{}_{\omega}(\mbox{$Q$-$\BTrans$})} &=& \emptyset\\
\delay_{\Id^{}_{\omega}(\mbox{$Q$-$\BTrans$})}(I) &=& \{(\sigma,\sigma)u  \mid  \sigma \in {\Store}, u \in I\} 
                              \cup\ \{(\sigma,\sigma)| \sigma \in {\Store}\}
\end{array}
\]
(We skip over the small difference between the notion of an $L_{\Res}$-model and of an algebra satisfying equations.)%

We write $\Cpo$ and $\SL$ for, respectively, the category of $\omega$-cpos and the category of  $\omega$-cpos with all finite sups. For any poset $P$, its lifting $P_{\perp}$ is the poset obtained from $P$ by freely adjoining a least element $\perp$; its elements are $(0,x)$, for $x \in P$, and $\perp$, and they are ordered in the evident way. If $P$ has all sups of increasing $\omega$-chains, i.e., is an $\omega$-cpo (respectively has finite sups), so does $P_{\perp}$.
For any object $a$ of any given category, and any set $X$, we write $X \otimes a$ and $a^{X}$ for, respectively, the $X$-fold sum and product of $a$ 
with itself, assuming they exist. The category $\SL$ has countable biproducts, given by the usual cartesian product of posets, and it is convenient to identify $X\otimes L$ with $L^{X}$, for countable sets $X$.

The next theorem shows that the algebraic notion of resumptions can indeed be characterized in trace-based terms, specifically as ideals of basic $Q$-transition sequences.
 \begin{thm} \label{resumptions}
Viewed as an $L_{\Res}$-model,  $\Id^{}_{\omega}(\mbox{$Q$-$\BTrans$})$ is $T_{\Res}(\Id^{\uparrow}_{\omega}(Q))$. 
The unit 
\[(\eta_{T_{\Res}})_{\Id^{\uparrow}_{\omega}(Q)}\type \Id^{\uparrow}_{\omega}(Q) \rightarrow \Id^{}_{\omega}(\mbox{$Q$-$\BTrans$})\] 
is given by:
\[(\eta_{T_{\Res}})_{\Id^{\uparrow}_{\omega}(Q)}(I) = \{(\sigma, \sigma\ x) \mid \sigma \in \Store, x \in I \}  
\]
and, for any continuous  $f \type \Id^{\uparrow}_{\omega}(Q) \rightarrow \Id^{}_{\omega}(\mbox{$R$-$\BTrans$})$, its Kleisli extension  
\[f^{\dagger} \type 
                      \Id^{}_{\omega}(\mbox{$Q$-$\BTrans$}) 
                      \rightarrow \Id^{}_{\omega}(\mbox{$R$-$\BTrans$})\]
is given by:
\[\begin{array}{lll} f^{\dagger}(I) & =  & \{ u (\sigma, \tau) v \mid
               \exists  \sigma', x.\;
               u (\sigma, \sigma'\ x) \in I,  
               (\sigma', \tau) v \in f(x\!\!\downarrow)\}\\
 & &                 \cup\ \{ u \in I \mid u \mbox{\rm \ has no $Q$-transition}\}
 \end{array}\]
 \end{thm}
\proof

Models of $L_{\Res}$ in $\Cpo$ correspond to models of $L_{\mathrm{S}}$ in $\SL$ together with a morphism $\delay'\type L_{\perp} \rightarrow L$, where $L$  is the carrier of the model. (Such morphisms are equivalent to $\omega$-continuous maps on $L$ which preserve  binary sups, but not necessarily $\perp$.) The carrier $L$ of the model of $L_{\Res}$ is that of the model of $L_{\mathrm{S}}$ in $\SL$; it is necessarily an $\omega$-cpo with all finite lubs. The $L_{S}$ operations on $L$ become those of the model of $L_{\mathrm{S}}$ in $\SL$, and the map $\delay\type L \rightarrow L$ extends uniquely to a morphism on
$L_{\perp}$, obtaining the required map $\delay'$. This correspondence extends  straightforwardly to an equivalence of categories.

 So, as $\Id^{}_{\omega}(Q)$ is the free $\omega$-cpo with finite sups over the $\omega$-cpo $\Id^{\uparrow}_{\omega}(Q)$, we seek the free structure 
\[(L,(\update_{l,n})_L, (\lookup_{l})_L, d_{L})\]
over $\Id^{}_{\omega}(Q)$, consisting of a model $(L,(\update_{l,n})_L, (\lookup_{l})_L)$  of $L_{\mathrm{S}}$ in $\SL$ and a morphism $\delay'\type L_{\perp} \rightarrow L$.

By Theorem 1 of~\cite{PP1} the free algebra monad  for $L_{S}$ over $\SL$  is $T_{S} = (S \otimes -)^{S}$, 
where we abbreviate $\Store$ to $S$ (the theorem depends on the set of variables being finite). The definitions of the operations 
$(\update_{l,n})_{T_{S}(L)}$ and  $(\lookup_{l})_{T_{S}(L)}$ of an algebra $T_{S}(L)$ are given by Proposition 1 of~\cite{PP1}; the unit $(\eta_{T_{\mathrm{S}}})_{L}$ at $L$ is the canonical map $L \longrightarrow (S \otimes L)^{S}$.

So, by Corollary  2  of~\cite{HPP}, for any poset $Q$, 
$L$ is the solution of the following ``domain equation'' in $\SL$:
\begin{equation} \label{*} L \cong (S \otimes  (L_{\perp} + \Id^{}_{\omega}(Q)))^S
\end{equation}
by which we mean the initial  $\omega$-cpo with finite sups $L$ and map 
\[\alpha\type (S \otimes  (L_{\perp} + \Id^{}_{\omega}(Q)))^S \rightarrow L\]
(Such a map is necessarily an isomorphism.)

The morphism $(\update_{l,n})_L\type L \rightarrow L$ is
\[L \xrightarrow{\alpha^{-1}} T_{\mathrm{S}}(L_{\perp} + \Id_{\omega}(Q)) \xrightarrow{(\update_{l,n})_{T_{S}(L_{\perp} + \Id_{\omega}(Q))}} T_{\mathrm{S}}(L_{\perp} + \Id_{\omega}(Q)) \xrightarrow{\alpha} L \]
 the morphism $(\lookup_{l})_L\type L^{\mathbb{N}} \rightarrow L$ is
\[L^{\mathbb{N}} 
                               \xrightarrow{(\alpha^{-1})^{\mathbb{N}}} 
  T_{\mathrm{S}}(L_{\perp} + \Id_{\omega}(Q))^{\mathbb{N}}
                               \xrightarrow{(\lookup_{l})_{T_{S}(L_{\perp} + \Id_{\omega}(Q))}} 
  T_{\mathrm{S}}(L_{\perp} + \Id_{\omega}(Q))
                                \xrightarrow{\alpha} 
 L \]
 the morphism $\delay'_L\type L_{\perp} \rightarrow L$ is
\[L_{\perp} \xrightarrow{\mathrm{inl}} L_{\perp} + \Id_{\omega}(Q) 
\xrightarrow{(\eta_{T_{\mathrm{S}}})_{(L_{\perp} + \Id_{\omega}(Q))}} 
T_{\mathrm{S}}(L_{\perp} + \Id_{\omega}(Q))
\xrightarrow{\alpha} L \]
and at $\Id^{\uparrow}(Q)$ the unit $\eta_{T_{\Res}}$ is
\[\Id^{\uparrow}_{\omega}(Q) 
      \hookrightarrow 
\Id^{}_{\omega}(Q) 
       \xrightarrow{\mathrm{inr}} 
L_{\perp} + \Id_{\omega}(Q) 
        \xrightarrow{(\eta_{ T_{\mathrm{S}}})_{(L_{\perp} + \Id_{\omega}(Q))}} 
T_{\mathrm{S}}(L_{\perp} + \Id_{\omega}(Q))
       \xrightarrow{\alpha} 
 L
\]
Now, since 
countable copowers and powers coincide in  $\SL$,  Equation~(\ref{*}) can be rewritten as:

\begin{equation} \label{**}
L  \cong  S \otimes (S \otimes (L_{\perp} + \Id^{}_{\omega}(Q)))
\end{equation} 
As $\Id_{\omega}\type\Pos \rightarrow \SL$ is a left adjoint, where $\Pos$ is the category of posets,
it preserves all colimits;  $\Id_{\omega}$  also commutes with lifting. So there is an isomorphism:
\[\beta\type \Id^{}_{\omega}(S \times (S \times (R_{\perp} + Q))) \cong S \otimes (S \otimes (\Id^{}_{\omega}(R)_{\perp} + \Id^{}_{\omega}(Q))) \]
for any poset $R$. So, again using that $\Id_{\omega}$ preserves all colimits, we can solve Equation~(\ref{**}) by first solving the equation:
\[R \cong S \times (S \times (R_{\perp} + Q))
\]
 in the category $\Pos$, and then applying $\Id_{\omega}$.
To do that, one takes $R$ to be the least set such that  
\[R = S \times (S \times (R_{\perp} + Q))\] 
and then imposes the evident inductively defined partial order on it. The solution of Equation~(\ref{**}) is then given by taking 
$L = \Id_{\omega}(R)$ 
and 
$\alpha = \beta^{-1}$.

We now have an expression of $L$ as $ \Id_{\omega}(R)$, as well as definitions of $(\update_{l,n})_L$,  $(\lookup_{l})_L$,
$\delay_L$,  and the unit. So, given the initial discussion above, we see that  $L$ forms the free model of $L_{\Res}$ over 
$\Id^{\uparrow}_{\omega}(R)$ with unit:
\[(\eta_{\Res})_{\Id^{\uparrow}_{\omega}(R)} (I) = \{(\sigma, (\sigma,\mathrm{inr}(x))) \mid  x \in I \}\]
and with operations:
\[\begin{array}{lcl} 
  (\update_{l,n})_{L}(I) & =  & \{(\sigma, (\sigma',u))\mid (\sigma[l \mapsto n], (\sigma',u))  \in I \}\\
  (\lookup_{l})_{L}((I_{n})_{n})  &=&  \{(\sigma, (\sigma',u)) \in I_{n}\mid n \in \mathbb{N}, \sigma(l) = n \}\\
  I \cup_{L} J &=& I \cup J\\
  \Omega_{L} &=& \emptyset\\
  
 \delay_L(I)  & = &  
        \{(\sigma, (\sigma,\mathrm{inl}(0,u))) \mid \sigma \in S, u \in I \} 
          \;  \cup \;  \{(\sigma, (\sigma, \perp)) \mid \sigma \in S \})

\end{array}\]
There is an evident isomorphism of partial orders $\theta_{\Res}\type R \cong \mbox{$Q$-$\BTrans$}$, given recursively by:
\[\begin{array}{lcl}
\theta_{\Res}((\sigma, (\sigma', \mathrm{inl}((0,u))))) & = &(\sigma,\sigma')\theta_{\Res}(u)\\
\theta_{\Res}((\sigma, (\sigma', \mathrm{inl}(\perp)))) & = &(\sigma,\sigma')\\
\theta_{\Res}((\sigma, (\sigma', \mathrm{inr}(x)))) & = &(\sigma,\sigma' x)\\
\end{array}\]
This induces an isomorphism $\Id^{}_{\omega}(R) \cong \Id^{}_{\omega}(\mbox{$Q$-$\BTrans$})$ of $\omega$-cpos, and so the free such model is also carried by $\Id^{}_{\omega}(\mbox{$Q$-$\BTrans$})$. Using this, and the above definitions of the operations and unit for  $\Id^{}_{\omega}(R)$,
one then verifies that the operations and unit for $\Id^{}_{\omega}(\mbox{$Q$-$\BTrans$})$ are  as required. 

As regards the formula for the Kleisli extension, that $f^{\dagger}(\eta_{T_{\Res}})_{\Id^{\uparrow}_{\omega}(Q)} = f$ is evident and that the purported extension is a morphism of models of $L_{\Res}$ is a calculation.
\qed

 One can go further and obtain a closely related, if less elementary, picture of $T_{\Res}(P)$ for an arbitrary $\omega$-cpo $P$: one needs a notion of ideal that takes the $\omega$-sups of $P$ into account.

\subsection{Asynchronous Processes} \label{Threads}

One might hope that $\PThread$ can be understood as an $\omega$-cpo of resumptions, and, indeed,  basic $\{\donem\}$-transition sequences and proper pure non-empty transition sequences are very similar. Define a map 
$\theta_{\PThread}\type\mbox{$\{\donem\}$-$\BTrans$} \rightarrow \PPSeq $ by:
\[\begin{array}{lcll}
\theta_{\PThread}(u(\sigma,\sigma' \ \donem)) & = & u(\sigma,\sigma') \donem &\\
\theta_{\PThread}(u) & = & u & (\mbox{if $u$ does not contain $\donem$})
\end{array}\]
Unfortunately, while $\theta_{\PThread}$ is a monotonic 
bijection, it is not an isomorphism of partial orders, as
$u(\sigma,\sigma')  \prefix u(\sigma,\sigma')\donem$ but 
$u(\sigma,\sigma')  {\not\leq_{\{\donem\}}} u(\sigma,\sigma'\ \donem)$.

There is a related programming language phenomenon. Denotationally, we have the inclusion: 
\[\dsem{(\fork{(\yield;\block)});C} \subseteq \dsem{(\fork \unit);C}\]
but not the inclusion:
\[\dsem{\yield;\block} \subseteq \dsem{\unit}\]
As in the proof of the full-abstraction theorem, one can distinguish $\dsem{\yield;\block}$ from $\dsem{\unit}$ using a sequential context; however,  this context is not available when the command is within an $\forkn$.

To solve this difficulty we  take the theory of asynchronous threads $L_{\PThread}$ to be $L_{\Res}$
extended by a new constant $\halt$ and the equation:
\[\delay(\Omega) \leq \halt\]
We can turn $\PThread_{}$ into a model of
$L_{\PThread}$ by defining operations  as follows:
\[\begin{array}{r@{\ }c@{\ }l}
(\update_{l,n})_{\PThread}(P)  &=& {
\{ (\sigma,\sigma')u \mid  (\sigma[l \mapsto n],\sigma')u \in P \} } \cup \{\varepsilon\}\\ 
(\lookup_l)_{\PThread} ((P_{n})_{n})  &=&  {
\bigcup_n \{(\sigma,\sigma')u \in P_{n} \mid  \sigma(l) = n\}} \cup \{\varepsilon\}\\
P \cup_{\PThread} Q &=& P \cup Q\\
\Omega_{\PThread} &=& \{\emptyseq\}\\
\delay_{\PThread}(P) &=& {\{(\sigma,\sigma)u \mid \sigma \in {\Store}, u \in P\}} \cup \{\varepsilon\}\\
\halt_{\PThread} &=& \prefixclosure{\{(\sigma,\sigma) \donem \mid \sigma \in \Store\}}
\end{array}
\]
Note that $\halt_{\PThread} = \clean{\dsem{\unit}}$.

We write $T_{\PThread}$  for the monad associated to the theory $\PThread$.
The next theorem shows that the variant theory $L_{\PThread}$ indeed captures $\PThread$. First we need some notation. 
\begin{enumerate}[$\bullet$]
	\item We define a unary derived operation $a_{l,m,k}$, for $l \in \Vars$ and $m,n \in \Value$ by:
	 \[a_{l,m,k}(x) \equiv_{\mathrm{def}} \lookup_l(( t_{m'})_{m'})\] 
where:
\[t_{m'}  \equiv_{\mathrm{def}}  \left \{\begin{array}{ll}
                       \update_{l,k}(x)   &   \mbox{(if $m' = m$)}\\
                       \Omega  &   \mbox{(otherwise)}
                        \end{array}
                                       \right.\]
	\item We define a unary derived operation $a_{\sigma,\sigma'}$, for $\sigma,\sigma' \in \Store$ by:
	\[a_{\sigma,\sigma'}(x) \equiv_{\mathrm{def}}  
	a_{l_1,\sigma(l_1),\sigma'(l_1)}(\ldots a_{l_n, \sigma(l_n),\sigma'(l_n)}(x)\ldots)\]
where $l_1,\ldots,l_n$ is an enumeration of $\Vars$.
	\item For every sequence of plain 
	transitions $u = (\sigma_1,\sigma'_1)\ldots (\sigma_n,\sigma'_n)$ we define a unary derived operation $a_u$ by:
\[a_{u}(x)  \equiv_{\mathrm{def}} 
           a_{\sigma_1,\sigma'_1}(d(\ldots a_{\sigma_n,\sigma_n'}(d(x))\ldots))\]
  \item For every sequence of plain transitions $u$ and $\sigma, \sigma' \in \Store$, we define two constants $\overline{u}$ and 
  $\overline{u(\sigma, \sigma')\donem}$ by:
  \[\overline{u} \equiv_{\mathrm{def}} a_{u}(\Omega) \qquad \mbox{and} \qquad \overline{u(\sigma, \sigma')\donem} \equiv_{\mathrm{def}} a_{u}(\halt)\]

\end{enumerate}
Note that $\overline{u}_{\PThread} =  \overline{u}_{\Id^{}_{\omega}(Q)} = u\downarrow$, where, for example,  $\overline{u}_{\PThread}$ is the interpretation of $\overline{u}$ in $\PThread$; further $\overline{u(\sigma, \sigma')\donem}_{\PThread} = u(\sigma, \sigma')\donem\downarrow$.
Below we may confuse a constant or operation with its interpretation in a specific algebra $A$, e.g., writing $\overline{u}$ or $a_{u}$  rather than  
$\overline{u}_{A}$ or $(a_{u})_A$, provided that the intended algebra can be understood from the context.

\begin{thm} \label{threads}
$\PThread_{}$ is the initial $L_{\PThread}$-model, i.e., it is $T_{\PThread}(0)$.
\end{thm}
\proof  We begin by examining the connection between  $\Id^{}_{\omega}(\{\donem\})$ and $\PThread$. By Theorem~\ref{resumptions},  $\Id^{}_{\omega}(\{\donem\})$ is the free model of $L_{\Res}$ over $\{\donem\}$.  So $f\type \{\donem\} \rightarrow \PThread$ has a unique extension to a morphism $f^{\dagger}\type \Id^{}_{\omega}(\{\donem\}) \rightarrow \PThread$ of $L_{\Res}$-models, where $f(\donem) \eqdef \halt_{\PThread}$. We now show that:
\[f^{\dagger}(I) = \{\theta_{\PThread}(u)\mid u \in I\}\downarrow\]
from which it follows that $f^{\dagger}$ is onto.
It is enough to show that $f^{\dagger}(u\downarrow) = \theta_{\PThread}(u)\downarrow$, which holds as, for any $u$ not containing $\donem$, we calculate that
\[f^{\dagger}(u\downarrow)  = f^{\dagger}(\overline{u})_{\Id^{}_{\omega}(\{\donem\})})  =  (\overline{u})_{\PThread} = u\downarrow = \theta_{\PThread}(u)\downarrow\]
and that
\[ \begin{array}{lclcl}
f^{\dagger}(u(\sigma,\sigma' \ \donem)\downarrow)
  & = &  f^{\dagger}((a_{u})_{\Id^{}_{\omega}(\{\donem\})}(\eta(\donem))) 
& = &  (a_{u})_{\PThread}(f^{\dagger}(\eta(\donem)))\\
 & = &   (a_{u})_{\PThread}(\halt_{\PThread})  
 & = &  u(\sigma,\sigma' ) \donem\downarrow \\
 & =  & \theta_{\PThread}(u(\sigma,\sigma' \ \donem))\downarrow
 \end{array}\]
where, in both cases, the second equality holds as $f^{\dagger}$ is a morphism  of $L_{\Res}$-models.

Let $L$ be a model of $L_{\PThread}$. We have to show there is a unique morphism $h\type \PThread \rightarrow L$. For uniqueness, let $h,h'$ be such morphisms.  Then both ${f^{\dagger}}\comp{h}$ and ${f^{\dagger}}\comp{h'}$ are morphisms of $L_{\Res}$ models from $\Id^{}_{\omega}(\{\donem\})$ to $L$, extending the map $\donem \mapsto \halt_{L}$. So, as there is only one such map,  ${f^{\dagger}}\comp{h} = {f^{\dagger}}\comp{h'}$, and therefore, as $f^{\dagger}$ is onto,  $h = h'$, as required.

For existence, define the map $\theta\type \PPSeq \rightarrow L$ by: $\theta(u) = (a_{u})_{L}$. Using the fact that $L$ is a model of $\PThread$, particularly the axiom $\delay(\Omega) \leq \halt$, one has that $\theta$ is monotonic. One can then define a continuous map $h \type \PThread \rightarrow L$ by:
\[h(I) = \bigvee_{u \in I} \theta(u)\]
with the sup on the right existing as $I$ is denumerable.  Let $g$ be  the unique morphism of $L_{\Res}$ models from $\Id^{}_{\omega}(\{\donem\})$ to $L$, extending the map $\donem \mapsto \halt_{L}$. 

We have that $h\comp f^{\dagger}= g$, as, for any $u$ not containing $\donem$, we may calculate that:
\[h(f^{\dagger}(u\downarrow))  =   h(u\downarrow) = \theta(u) = a_{u} = g(a_{u}) =  g(u)\]
and that
\[\begin{array}{lclcl}
h(f^{\dagger}(u(\sigma,\sigma' \ \donem)\downarrow))
& =  & h(a_{u}(\halt))  
& =  & h(u(\sigma,\sigma' ) \donem\downarrow) \\
& =  & \theta(u(\sigma,\sigma' ) \donem) 
 & =  & a_{u(\sigma,\sigma' ) \donem} \\
 & =  & a_{u(\sigma,\sigma' )}( \halt) 
 & =  & g(a_{u(\sigma,\sigma' )}( \eta(\donem))) \\
 & =  & g(u(\sigma,\sigma' \, \donem))\end{array}\]
As $h\comp f^{\dagger} = g$, and $f^{\dagger}$ and $g$ are morphisms of $L_{\Res}$ models, and $f^{\dagger}$ is onto, $h$ is automatically a morphism of $L_{\Res}$ models. For example, for the preservation of $\delay$, given $I\in \PThread$, choose $J \in \Id^{}_{\omega}(\{\donem\})$ such that  $f^{\dagger}(J) = I$ and calculate that: 
\[\begin{array}{lclcl}
h(\delay_{\PThread}(I)) 
& =  &  h(\delay_{\PThread}(f^{\dagger}(J))) \\
&  =  &  h(f^{\dagger}(\delay_{\Id^{}_{\omega}(\{\donem\})}(J)))
&  = &  g(\delay_{\Id^{}_{\omega}(\{\donem\})}(J))\\
& = & \delay_{L}(g(J)) 
&  = & \delay_{L}(h(f^{\dagger}(J))) \\
& = & \delay_{L}(h(I))
\end{array}\]
Further,
 $h$ preserves $\halt$ as $h(\halt_{\PThread})$ = $\theta(\halt_{\PThread}) = \halt_{L}$. We therefore have that $h$ is a morphism of $L_{\PThread}$-models, which concludes the proof.
\qed

One can  go on and obtain a general view of the monad $T_{\PThread}$ using a suitable notion of (proper) pure $Q$-transition sequences. However we omit the details as they are not needed for an account of processes.
 
There is another possible proof of Theorem~\ref{threads} along the lines of that of  Theorem~\ref{resumptions}.
First one notes that to have a model of $L_{\PThread}$ in $\Cpo$ is to have a model of $L_{\mathrm{S}}$ in $\SL$, with carrier $L$, say, together with a morphism $\delay\type L_{\perp} \rightarrow L$ and an element $\halt \in L$ such that $\delay(\Omega) \leq \halt$. It is not hard to see that to have such a morphism and element is to have a morphism $(L + \Id_{\omega}(\mathbbm{1}))_{\perp} \rightarrow L$, where $\mathbbm{1}$ is the one-point partial order.

One then sees that the carrier of the initial such model is given by the solution of the domain equation:
\[ L \cong (S \otimes  (L + \Id_{\omega}(\mathbbm{1}))_{\perp} )^S\]
and that that can be solved by first solving the corresponding equation
\[ R \cong S \times  (S \times (R + \mathbbm{1})_{\perp} )\]
in $\Pos$ and then setting $L = \Id_{\omega}(R)$. 
The rest of  the proof proceeds  as expected. 

Equally, there should be an elementary proof of Theorem~\ref{resumptions}, which, like that of Theorem~\ref{threads}, makes use of definability. The more conceptual proofs have the advantage of showing, via domain equations, the origins of the two kinds of transition sequences and their ordering.

\subsection{Processes} \label{Processes}

We turn to our algebraic account of $\Proc$. The signature of our theory of processes, 
$L_{\Proc}$, is that for $L_{\Res}$ together with two families of unary operation symbols $\async_{P}$ and $\rsync_{P}$, where $P$ is in $\PThread$.
The first of these corresponds to the function of the same name defined above, but restricted to asynchronous threads. The second corresponds to a slightly different version of $\async$ in which the first action is that of the thread spun off, rather than that of the active command. We often find it convenient to write 
$\async_P{t}$ and $\rsync_P{t}$ as, respectively, $P\rlaction t$ and $P \llaction t$,  thinking of them as right and left shuffles.

We begin with a theory $L_{\Spawn}$ for $\async$ and $\rsync$ which involves the other operations. The first group of equations for $L_{\Spawn}$ concerns commutation with $\cup$:
\[
\begin{array}{rcl}
(P\cup_{\PThread}  P') \rlaction x &=& (P  \rlaction x) \cup (P' \rlaction x) \\
P  \rlaction (x \cup y) &=& (P  \rlaction x) \cup (P  \rlaction y)\\
(P \cup_{\PThread} P') \llaction x &=&  (P \llaction x) \cup  (P' \llaction x)\\
 P \llaction (x \cup y) &=& (P \llaction x) \cup (P  \llaction y)
 \end{array}
\]

The second group of equations concerns the interaction of $\async$ with the other operations of $L_{\Proc}$ (except for 
$\llaction$):
\[
\begin{array}{rcl}
P  \rlaction  \update_{l,n}(x)  &=&   \update_{l,n}( P  \rlaction x)\\
P  \rlaction \lookup_l((x_n)_n)  &=&  \lookup_l((P  \rlaction x_n)_n)\\
P \rlaction \Omega &=&  \Omega\\
P \rlaction \delay(x) &=&  \delay(P \bowtie x)\\
P  \rlaction (P'  \rlaction x) &=& (P \bowtie P')  \rlaction x
\end{array}
\]
where we write $P\bowtie x$ for the ``left action'' $(P \rlaction x) \cup (P\llaction x)$.
The first three state that $P\rlaction - $ commutes with another operation; the next concerns the interaction of $\async$ with suspension and brings in $\rsync$; the last reduces two occurrences of $\async$ to one. 
The third, and last,  group of equations is for the interaction of $\rsync$ with the other operations of $L_{\PThread}$:
\[
\begin{array}{rcl}
  (\update_{l,n})_{\PThread}(P) \llaction x &=& \update_{l,n}(  P \llaction x)\\
  (\lookup_l)_{\PThread}((P_{n} )_{n}) \llaction x &=& \lookup_l((P_{n} \llaction x)_n)\\
  \Omega_{\PThread} \llaction x &=& \Omega\\
  \delay_{\PThread}(P) \llaction x&=& \delay(P \bowtie x)\\
  \halt_{\PThread} \llaction x &=& \delay(x)
\end{array}
\]
The first three assert that $- \llaction x$ acts homomorphically with respect to an operation; the next concerns the interaction with suspension; and the last concerns what happens when asynchronous threads halt.
Finally we add an inequation:
\[\Omega_{\PThread} \rlaction x  \leq x \]

We  take the equations of $L_{\Proc}$ to be those of $L_{\Spawn}$, i.e., the equations are the ones just given for 
$\async$ and $\rsync$, together with those of $L_{\Res}$.
One would naturally have expected  $L_{\Proc}$ also to have 
an equation with left-hand side $P \rlaction (P' \llaction x)$; indeed, we could have added the equation:
\[P \rlaction (P' \llaction x)  = P'\llaction (P \bowtie x)\]
However this equation is redundant as it can be proved from the others using the algebraic induction principle of ``Computational Induction'' described in~\cite{DBLP:conf/lics/PlotkinP08}. (One proceeds by such an induction  on $P'$, with a  subinduction on $P$.)
The inequation is somewhat inelegant: a possible improvement would be  to use $\Thread$ instead rather than restricting to asynchronous threads. This would give the possibility of a version of $\halt$, to denote $\donem\!\downarrow$, such that the equations
\[\halt \rlaction x = \halt \llaction x =  x \]
held, making the inequation redundant.

\newcommand{\QTrans}{\mbox{$Q$-$\mathrm{Trans}$}}
\newcommand{\QProc}{\mbox{$Q$-$\mathrm{Proc}$}}

Let $T_{\Proc}$  be the monad associated to the theory $\Proc$. 
We now aim to give a picture of $T_{\Proc}(\Id^{\uparrow}_{\omega}(Q))$ like that we gave 
of $T_{\Res}(\Id^{\uparrow}_{\omega}(Q))$. 
Take the partial order $\QTrans$ of the \emph{$Q$-transition sequences} to be that of the basic $(Q \times \PSeq)$-transition sequences. Note that one can regard $Q$-transition sequences as elements of a kind of ``double thread'' in which the first thread returns a value together with a second (asynchronous) thread. 

We  show that $\QProc \eqdef \Id_{\omega}(\QTrans)$ carries the free model of $L_{\Proc}$ on  
$\Id^{\uparrow}_{\omega}(Q)$. We view $\QProc$ as a $L_{\Res}$-model as in Section~\ref{Resumptions}. 
In order  to give $\async$ and 
$\rsync$, we first mutually recursively define the incomplete right and left 
shuffles $u \rlaction v \mbox{ and } u \llaction v \mbox{ in } \QProc$ 
of a proper pure transition sequence $u$ with a $Q$-transition sequence
$v$, by:
\[\begin{array}{lclr}
	u\rlaction (\sigma,\sigma') & = & \prefixclosure{\{(\sigma,\sigma')u^{-}\}}&\\
	u\rlaction (\sigma,\sigma' (x,u')) & = & \{(\sigma,\sigma' (x,w))\mid w \in u
	 \bowtie u'\}\downarrow&\\
u \rlaction (\sigma,\sigma')v & = & \{(\sigma,\sigma')w\mid w \in u \bowtie v\}\downarrow & \quad(v\neq \varepsilon)
\end{array}\]
where, for any pure transition sequence $w$, $w^{-}$ is $w$ less any occurrence of $\donem$,
and writing $u \bowtie v$ for the incomplete shuffles
 $(u\llaction v) \, \cup \,  (u \rlaction v)$ of $u$ and $v$,
and:
\[\begin{array}{lcl}
	\varepsilon \llaction v& = & \emptyset\\
	 (\sigma,\sigma')\donem \llaction v& = &\{(\sigma,\sigma') v\}\downarrow\\
	 (\sigma,\sigma')u \llaction v& = & \{(\sigma,\sigma')w\mid w \in u\bowtie v \}\downarrow
\end{array}\]
where, in the last line, $u$ is required to be proper.  
(Recall that an incomplete shuffle of two sequences is a shuffle of two of their prefixes, equivalently a prefix of a shuffle of them.)
Both $\rlaction$ and $\llaction$ are monotonic
 operations.

Then, for  $P \in \PThread$ and $I \in \QProc$, we put:
\[(\async_{\Proc})_P(I) = \!\!\!\bigcup_{u \in P,\, v\in I} u\rlaction v\]
\[(\rsync_{\Proc})_P(I) =  \!\!\!\bigcup_{u \in P, \,v\in I} u \llaction v \;\cup \; \{u^{-}\mid u \in P, u \neq \varepsilon\}
\]
If  $I$ is not empty we have:
\[(\rsync_{\Proc})_P(I) =  \!\!\!\bigcup_{u \in P, \,v\in I} u \llaction v
\]
With these additional operations, $\QProc$ is a model of $L_{{\Proc}}$.

In the following we make use of the notation introduced in Section~\ref{Threads}.
\begin{lem}\label{main_proc_lemma1} For any proper pure transition sequence $u$, the equation $ u\!\!\downarrow \! \llaction \;\Omega= \overline{u^{-}}$ is provable in $L_{\Proc}$.
\end{lem}
\proof
 The proof is by induction on the length of $u$.
In the case where $u = \varepsilon$, we have  $u\!\downarrow = \Omega_{\PThread}$, and in the equational theory we have $\Omega_{\PThread} \llaction \Omega = \Omega$, as required.

In the case where  $u = (\sigma,\sigma')$, we have 
$u\!\downarrow  = a_{\sigma,\sigma'}(\delay\Omega_{\PThread})$, and in the equational theory, we have:
	\[\begin{array}{lcl}
       a_{\sigma,\sigma'}(\delay\Omega_{\PThread})\llaction \Omega	& = &
              a_{\sigma,\sigma'}( \delay\Omega_{\PThread} \llaction \Omega	)\\
   & = & a_{\sigma,\sigma'}(\delay(\Omega_{\PThread} \llaction \Omega) \cup \delay(\Omega_{\PThread}\rlaction \Omega))\\
   & = & a_{\sigma,\sigma'}(\delay\Omega)
	\end{array}\]

	In the case where  $u = (\sigma,\sigma')\donem$, we have 
	$u\!\downarrow  = a_{\sigma,\sigma'}(\halt)$, and  in the equational theory, we have:	
	\[\begin{array}{lcl}
		 a_{\sigma,\sigma'}(\halt) \llaction \Omega& = & a_{\sigma,\sigma'}( \halt \llaction  \Omega) \\
		& = & a_{\sigma,\sigma'}(\delay\Omega)
	\end{array}\] 
	
Finally, in the case where  $u = (\sigma,\sigma')v$, with $v$  a proper pure transition sequence, we have	$u\!\downarrow  = a_{\sigma,\sigma'}(\delay (v\!\downarrow))$, and  in the equational theory, we have:
\[\begin{array}{lcl}
  a_{\sigma,\sigma'}(\delay (v\!\downarrow))\llaction \Omega & = & a_{\sigma,\sigma'}(\delay (v\!\downarrow) \llaction \Omega)\\
 & = & a_{\sigma,\sigma'}(\delay( (v\!\downarrow) \llaction \Omega) \cup \delay((v\!\downarrow)\rlaction \Omega))\\ 		
 & = & a_{\sigma,\sigma'}(\delay( (v\!\downarrow) \llaction \Omega)) \\
& = & a_{\sigma,\sigma'}(\delay(\overline{v^{-}}))\\
& = & \overline{u^{-}}
 \end{array}\]
using the induction hypothesis in the next-to-last step.
\qed

Our main algebraic theorem characterizes  free models of a natural equational theory for resumptions with thread-spawning in terms of a kind of double-thread. 
 \begin{thm}\label{thm:view} \label{processes}
Viewed as an $L_{\Proc}$-model, $\Id^{}_{\omega}(\QTrans)$ is the free model over 
$\Id^{\uparrow}_{\omega}(Q)$. 
The unit $(\eta_{T_{\Proc}})_{\Id^{\uparrow}_{\omega}(Q)}\type \Id^{\uparrow}_{\omega}(Q) \rightarrow \Id^{}_{\omega}(\QTrans)$ is given by:
\[(\eta_{T_{\Proc}})_{\Id^{\uparrow}_{\omega}(Q)}(I) = \{(\sigma, \sigma\ (x,\donem)) \mid x \in I \}\downarrow\]
and, for any continuous  $f \type \Id^{\uparrow}_{\omega}(Q) \rightarrow \Id^{}_{\omega}(\mbox{$R$-$\Trans$})$, its Kleisli extension 
\[f^{\dagger} \type 
                      \Id^{}_{\omega}(\QTrans) 
                      \rightarrow \Id^{}_{\omega}(\mbox{$R$-$\Trans$})\]
is given by:
\[\begin{array}{rcl} f^{\dagger}(I) & =  & \{ u (\sigma, \tau) v \mid
               \exists  \sigma', x.\;
               u (\sigma, \sigma'\ (x,\donem)) \in I, \\
            & &   
               \quad\quad\quad\quad\quad\quad(\sigma', \tau) v \in  f(x\!\!\downarrow)\}\\ 
  &&        \cup \{ u (\sigma, \tau) v \mid
               \exists  \sigma', x, w \neq \donem.\;
               u (\sigma, \sigma'\ (x,w)) \in I, \\
            & &   
               \quad\quad\quad\quad\quad\quad(\sigma', \tau) v \in w\downarrow\rlaction f(x\!\!\downarrow)\}\\             
 & &                 \cup\ \{ u \in I \mid  u \mbox{\rm \ has no $(Q \times  \PSeq)$ transition}\}
 \end{array}\]

\end{thm}
\proof
	To show that $\Id^{}_{\omega}(\QTrans)$ is the free algebra over 
	$\Id^{\uparrow}_{\omega}(Q)$ with unit  
	as above, we must show that for any $L_{\Proc}$-model $A$ and any continuous function $f\type \Id^{\uparrow}_{\omega}(Q) \rightarrow A$ there is a  unique morphism $h\type\Id^{}_{\omega}(\QTrans) \rightarrow A$ of models of $L_{\Proc}$ such that  the following diagram commutes:
	\begin{diagram}
	\Id^{\uparrow}_{\omega}(Q)\\
	 \dTo^{\;{(\eta_{T_{\Proc}})_{\Id^{\uparrow}_{\omega}(Q)}}} & \SE_{}{\quad \; f}\\
	\Id^{}_{\omega}(\QTrans) & \rTo^{h} & A
	\end{diagram}
	We begin by showing uniqueness. To that end, 
	fix $A$ and  $f$, and let $h$ be a morphism such that the diagram commutes.
Define $g\type \Id^{\uparrow}_{\omega}(Q\times \PSeq) \longrightarrow A$ by putting:
\[g((x,u)\!\downarrow) = \left\{\begin{array}{ll}
                                                            f(x\!\downarrow)  &  (\mbox{if u = $\donem$})\\
                                                            u\!\downarrow\rlaction_A f(x\!\downarrow) & (\mbox{otherwise})\\
                                                    \end{array}   
                                           \right.
\]
This is a good definition, with monotonicity being established using the inequation for $\rlaction$. 
We have $f = g\alpha$ and 
$(\eta_{T_{\Proc}})_{\Id^{\uparrow}_{\omega}(Q)} = (\eta_{T_{\Res}})_{\Id^{\uparrow}_{\omega}(Q \times \PSeq)}\alpha$ where $\alpha\type \Id^{\uparrow}_{\omega}(Q) \rightarrow \Id^{\uparrow}_{\omega}(Q\times \PSeq)$ is defined by setting $\alpha(x\!\downarrow) = (x,\donem)\!\downarrow$.

We then have that the following diagram commutes:
\begin{diagram}
	\Id^{\uparrow}_{\omega}(Q\times \PSeq)\\
	\dTo^{\; (\eta_{T_{\Res}})_{\Id^{\uparrow}_{\omega}(Q \times \PSeq)}} & \SE_{}{\; g}\\
	\Id^{}_{\omega}(\mbox{$(Q\times \PSeq)$-$\BTrans$}) & \rTo^{h} & A
	\end{diagram}
as we may we calculate, for $u = \donem$, that:
\[\begin{array}{lcl}
	h( (\eta_{T_{\Res}})_{\Id^{\uparrow}_{\omega}(Q \times \PSeq)}((x,u)\downarrow)) & = & h(\eta(x\!\downarrow))\\
		& = & f(x\!\downarrow)\\
		&=& g((x,u)\!\downarrow)
\end{array}\]
and, for $u \neq \donem$, that:
\[\begin{array}{lcl}
	h( (\eta_{T_{\Res}})_{\Id^{\uparrow}_{\omega}(Q \times \PSeq)}((x,u)\downarrow)) & = & h(\{(\sigma,\sigma (x,u)) \mid \sigma \in \Store\}\downarrow)\\
		& = & h(u\downarrow\rlaction\{(\sigma,\sigma (x,\donem)) \mid \sigma \in \Store\}\downarrow)\\
		& = & h(u\downarrow\rlaction (\eta_{T_{\Proc}})_{\Id^{\uparrow}_{\omega}(Q)}(x\downarrow))\\
		& = & u\downarrow\rlaction_{A} h((\eta_{T_{\Proc}})_{\Id^{\uparrow}_{\omega}(Q)}(x\downarrow))\\
		& = & u\downarrow\rlaction_{A} f(x\downarrow)\\
		&=& g((x,u)\!\downarrow)
\end{array}\]

This is enough to show uniqueness, as if $h(\eta_{T_{\Proc}})_{\Id^{\uparrow}_{\omega}(Q)} = h'(\eta_{T_{\Proc}})_{\Id^{\uparrow}_{\omega}(Q)} = f$, for two such morphisms $h$ and $h'$, then 
$h(\eta_{T_{\Res}})_{\Id^{\uparrow}_{\omega}(Q \times \PSeq)} = h'(\eta_{T_{\Res}})_{\Id^{\uparrow}_{\omega}(Q \times \PSeq)} = g$, and so $h = h'$, as $h$ and $h'$ are morphisms of models of $L_{\Res}$ (being morphisms of models of $L_{\Proc}$).
 
For existence  we are again given $A$ and $f$ and  wish to construct a suitable $h$. To that end, with $g$ and $\alpha$ as before, take $h$ to be the $T_{\Res}$-extension of $g$. Then we have $h(\eta_{T_{\Proc}})_{\Id^{\uparrow}_{\omega}(Q)} = h(\eta_{T_{\Res}})_{\Id^{\uparrow}_{\omega}(Q \times \PSeq)}\alpha = g\alpha = f$ and so it remains to prove that $h$ preserves $\async$ and $\rsync$.

As regards the preservation of $\async$, since it is continuous, preserves $\cup$ in each argument, and is strict in its second argument, it suffices to establish preservation for
individual transition sequences.
That is,  it suffices to show, for all  proper pure transition sequences $u$ and all $v$ in $\QTrans$, that:
\[h(u\rlaction v) = u\rlaction_Ah(v) \]
where here, and below, we omit $\downarrow$'s, writing, e.g., $u$ and $v$ rather than $u\!\!\downarrow$ and $v\!\!\downarrow$. 

As regards the preservation of $\rsync$, since it is continuous and preserves $\cup$ in each argument, it suffices  to show, for all  proper pure transition sequences $u$ and  all $v$ in $\QTrans$ that:
\[h(u \llaction v) = u \llaction_{A} h(v)\]
and:
\[h(u \llaction \Omega) = u \llaction_{A} h(\Omega)\]

For the last of these three equations, as $h(\Omega) = \Omega$, using Lemma~\ref{main_proc_lemma1},
 we see that is enough to show that
$h(\overline{u^{-}}) = \overline{u^{-}}$, and this holds as $h$ is a homomorphism of models of $L_{\Res}$.

The proof of the first two equations is a simultaneous induction on the sum of the lengths of $u$ and $v$, invoking $L_{\Proc}$ equations on $A$ as necessary.
We begin with the first equation. In the first case, we consider $v = (\sigma,\sigma')$.  Here, on the one hand, we have:
\[h(u\rlaction(\sigma,\sigma')) = h((\sigma,\sigma')u^{-}) = h(\overline{(\sigma,\sigma')u^{-}}) = \overline{(\sigma,\sigma')u^{-}}\]
using the fact that $h$ is a homomorphism for the last equality,
and, on the other, we have:
\[\begin{array}{lcl}
	u\rlaction_A h((\sigma,\sigma')) 
	& = & u\rlaction_A 	h(a_{\sigma,\sigma'}({\delay}\Omega))  \\
	& = & 
	u\rlaction_A(a_{\sigma,\sigma'}({\delay}\Omega)) \\
	& = & a_{\sigma,\sigma'}(u\rlaction_A {\delay}\Omega)  \\
	& = & a_{\sigma,\sigma'}(\delay(u\rlaction_A \Omega) \cup \delay(u \llaction_{A} \Omega)) \\
	& = & a_{\sigma,\sigma'}(\delay(\overline{u^{-}})) \quad\quad\mbox{(by Lemma~\ref{main_proc_lemma1})}  \\
	& = & \overline{(\sigma,\sigma')u^{-}}
	
\end{array}\]
For the next case we consider $v = (\sigma,\sigma' (x,u'))$. Here, on the one hand we have:
\[\begin{array}{lclcl}
	h(u\rlaction v) 
	& = & h(\{ (\sigma,\sigma' (x,u''))\mid u'' \in u \bowtie u'\}) \\
	& = & \bigcup_{u'' \in u \bowtie u'} h((\sigma,\sigma'(x,u'')))\\
	& = & \bigcup_{u'' \in u \bowtie u'} 
	       a_{\sigma,\sigma'}(h((\eta_{T_{\Res}})_{\Id^{\uparrow}_{\omega}(Q \times \PSeq)}(x,u'')))\\
	& = & \bigcup_{u'' \in u \bowtie u'} a_{\sigma,\sigma'}(u''\rlaction_A f(x))\\
	& = &  a_{\sigma,\sigma'}(\bigcup_{u'' \in u \bowtie u'}  (u''\rlaction_A f(x)))\\
	& = &  a_{\sigma,\sigma'}( (u\bowtie u') \rlaction_A f(x))
\end{array}\]
and, on the other hand, we have:
\[\begin{array}{lcl}
u\rlaction_A h(v)
	& = & u\rlaction_A h(a_{\sigma,\sigma'}((\eta_{T_{\Res}})_{\Id^{\uparrow}_{\omega}(Q \times \PSeq)}(x,u')))\\
	& = & a_{\sigma,\sigma'}(u\rlaction_A h((\eta_{T_{\Res}})_{\Id^{\uparrow}_{\omega}(Q \times \PSeq)}(x,u')))\\
         & = & a_{\sigma,\sigma'}(u\rlaction_A (u'\rlaction_A f(x)))
\end{array}\]
For the last case for the first equation we have $v = (\sigma,\sigma')v'$, with $v'$ in $\QTrans$, and we calculate:
\[\begin{array}{cllcl}
h(u\rlaction(\sigma,\sigma')v') 
 	& \!=\! & a_{\sigma,\sigma'}(h(u\rlaction {\delay}(v')))
	& \!=\! & a_{\sigma,\sigma'}(h({\delay}(u\rlaction v' \cup u \llaction v')))\\
	& \!=\! & a_{\sigma,\sigma'}({\delay}(h(u\rlaction v') \cup h(u \llaction v')))
	& \!=\! & a_{\sigma,\sigma'}({\delay}(u\rlaction_A h(v') \cup u \llaction_{A} h(v')))\\
	& \!=\! & a_{\sigma,\sigma'}(u\rlaction_A ({\delay}(h(v'))))
	& \!=\! & u\rlaction_A a_{\sigma,\sigma'}({\delay}(h(v')))\\
	& \!=\! & u\rlaction_A h((\sigma,\sigma')v')
	&&
	\end{array}\]
applying the induction hypothesis in the second line.

Turning to the second equation, the first case we consider is where $u = \varepsilon$, and we have: 
\[\begin{array}{lclclclclcl}
h(\varepsilon \llaction v)  
	& = & h(\Omega \llaction v)
	& = & h(\Omega)
	& = & \Omega 
	& = & \Omega \llaction_{A} h(v)
	& = & \varepsilon \llaction_{A }h(v)
\end{array}\]
The second case is where $u = (\sigma,\sigma')\donem$ and we have:
\[\begin{array}{lclcl}
h((\sigma,\sigma')\donem \llaction v) 
	& = & h((\sigma,\sigma')v) 
	& = & a_{\sigma,\sigma'}({\delay}(hv))\\
	& = & a_{\sigma,\sigma'}(\halt \llaction_{A} h(v))
	& = & a_{\sigma,\sigma'}(\halt) \llaction_A h(v)\\
	& = & (\sigma,\sigma')\donem \llaction_A h(v)
\end{array}\]
The last case is where $u = (\sigma,\sigma')u'$, with  $u'$ a proper pure transition sequence, 
and we have:
\[\begin{array}{lclcl}
h((\sigma,\sigma') u' \llaction v)
	& = & h(a_{\sigma,\sigma'}(\delay(u')) \llaction v)
	& = & h(a_{\sigma,\sigma'}(\delay(u'  \bowtie v)))\\
	& = & a_{\sigma,\sigma'}(\delay(h(u'  \bowtie v)))
         & = & a_{\sigma,\sigma'}(\delay(u'  \bowtie_{A} h(v)))\\
         & = & a_{\sigma,\sigma'}(\delay(u')) \llaction_{A} h(v)
         & = & (\sigma,\sigma') u' \llaction_{A} h(v)
	\end{array}\]
applying the induction hypothesis to obtain the fourth equality.

Finally, the formula for the Kleisli extension follows from the construction of $h$, using the Kleisli formula of 
Theorem~\ref{resumptions}.
\qed

As in the case of resumptions, one can go further and obtain a closely related, if less elementary, picture of  $T_{\Proc}(P)$ for arbitrary~$P$. 

Note that the proof of Theorem~\ref{processes} is elementary, making use of definability in a similar way to the proof of Theorem~\ref{threads}. However, unlike in the cases of Theorems~\ref{resumptions} and \ref{threads}, we do not know any conceptual proof of Theorem~\ref{processes}. The difficulty is that the theory of processes $L_{\Proc}$, particularly the part concerning $\llaction$ and $\rlaction$, seems somewhat ad hoc, and is not built up in a standard way from  simpler theories. There is surely more to be understood here.

Nonetheless, with Theorem~\ref{processes} available, we are in a position to give our algebraic account of $\Proc$.
There is an isomorphism $\theta_{\Proc}\type \QTrans \rightarrow \TSeq  \! \setminus \! \{\varepsilon\}$, where $Q = \{\finalm\}$, sending 
$u = (\sigma_1,\sigma'_1)\ldots(\sigma_n,\sigma'_n)$ to itself and $u(\sigma,\sigma'\ (\finalm,v))$ to 
$u(\sigma,\sigma'\ \finalm) v$.  One then has an  isomorphism  of $\omega$-cpos ${\tilde\theta_{\Proc}} \type \Id_{\omega}(\QTrans) \cong \Proc$ 
given by: $\tilde\theta_{\Proc}(I) =  \theta_{\Proc}(I) \cup \{\varepsilon\}$. It follows that
$\Proc$ can be seen as the free model of $L_{\Proc}$ over the terminal $\omega$-cpo $\{\finalm\}$, as we now spell out. 
First, define the set of left shuffles  $u\llaction v$ of a pure transition sequence $u$ with a transition sequence $v$ by setting 
\[
 \emptyseq \llaction v = \{\emptyseq\}
\]
and
\[
 (\sigma, \sigma') u \llaction v = \{(\sigma, \sigma')w \mid w \in u \bowtie {v}\}
\] 
Then, we have:
\begin{cor} \label{Proc-algebra}
Equip $\Proc$ with the following operations:
\[
\begin{array}{rcl}
(\update_{x,n})_{\Proc}(P) &=& \{(\sigma,\tau)u \mid (\sigma[x \mapsto n],\tau)u \in P \} \cup \{\varepsilon\}\\
(\lookup_x)_{\Proc} ((P_{n})_{n}) &=& \bigcup_{n} \{(\sigma,\tau)u \in P_{n} \mid \sigma(x) = n\} \cup \{\varepsilon\} \\
P \cup_{\Proc} Q &=& P \cup Q\\
\Omega_{\Proc} &=& \{\varepsilon\}\\
\delay_{\Proc}(P) &=& \{(\sigma,\sigma)u  \mid  \sigma \in {\Store}, u \in P\} \cup \{\varepsilon\}\\
P \rlaction_{\Proc} Q & = & \async(P,Q)\\
P \llaction_{\Proc} Q & = &  \bigcup_{u \in P, v \in Q} u \llaction v
\end{array}
\]
(where $x$ ranges over $\fun{Vars}$). 

Then $\tilde\theta_{\Proc} \type \Id_{\omega}(\QTrans) \cong \Proc$ is an isomorphism of $L_{\Proc}$-models, and  $\Proc$ is the free model of $L_{\Proc}$ over $\{\finalm\}$, with unit  $(\eta_{\Proc})_{\{\finalm\}}\type \{\finalm\} \rightarrow \Proc$
given by: 
\[(\eta_{\Proc})_{\{\finalm\}}(\finalm) =  \{(\sigma, \sigma \,\finalm)\donem \mid  \sigma \in {\Store}\}\!\downarrow\]
The Kleisli extension of a map $f\type \{\finalm\} \rightarrow \Proc$ is given by:
\[f^{\dagger}(P) = P \comp f(\finalm)\]
\end{cor}
\proof
The proof is a calculation using Theorem~\ref{thm:view}. The following equations are useful:
\[\tilde \theta_{\Proc}(u \rlaction v) =  (u \rlaction \theta_{\Proc}(v))\downarrow\]
\[\tilde \theta_{\Proc}(u \llaction v) =  (u \llaction \theta_{\Proc}(v))\downarrow\]
where $u$ is a proper pure transition sequence 
and $v$ is a $\{\finalm\}$-transition sequence.

\qed
As we now see, the algebraic view also determines the semantics of our language.
This  achieves our aim of placing cooperative threads within the algebraic approach to effects, thereby justifying the previous, more ad hoc, account.

First, we have 
that $\dsem{\unit} = (\eta_{\Proc})_{\{\finalm\}}(\finalm)$ 
and 
that $P \comp Q = (\finalm \mapsto Q)^{\dagger}(P)$, 
so the Kleisli structure determines the semantics of $\unit$ and composition, just 
as one would expect from the monadic point of view. 

Next, the $\update$ and $\lookup$ operations, together with the assumed primitive natural number and boolean functions, determine the semantics of assignments, conditionals, and while loops. The operations are equivalent to two generic effects, of assignment and reading:
\[:= \; \, \type \Vars  \times \mathbb{N} \rightarrow \Proc \qquad !\type \Vars \rightarrow T_{\Proc}(\mathbb{N})\]
One can use the reading generic effect  to give the semantics of numerical expressions as elements of $T_{\Proc}(\mathbb{N})$; with that,  one can give the semantics of assignments, using the assignment generic effect, standard monadic means, and $\tilde\theta_{\Proc}$. Similarly, one can use the reading generic effect  to give the semantics of boolean expressions as elements of $T_{\Proc}(\mathbb{B})$, where $\mathbb{B} \eqdef \{\btrue, \bfalse\}$; with that one can give the semantics of conditionals and while loops, again using standard monadic means and $\tilde\theta_{\Proc}$ (as well as least fixed-points for while loops).
 
 Continuing, the $\delay$ operation is that of the algebra; and $\block$ is modeled by $\Omega_{\Proc}$. 
Finally, the semantics of spawning is determined by $\async$ together with the cleaning function \[-^c\type \Proc \rightarrow \PThread\] It turns out that the latter is also determined by algebraic means. 
Specifically, one can regard $\PThread$ as a model of $L_{\Res}$ as in Section~\ref{Threads} (so we ignore $\halt$) and then extend it to a model of $L_{\Proc}$ as follows. First for any proper pure transition sequences $u$ and $v$ we define $u \rlaction v \in \PThread$ inductively on $v$ by:
\[\begin{array}{lcl}
	u \rlaction \varepsilon & = & \{\varepsilon \}\\
	u \rlaction (\sigma,\sigma')\donem & = &\{(\sigma,\sigma') u\}\downarrow\\
	u \rlaction  (\sigma,\sigma')v & = & \{(\sigma,\sigma')w\mid w \in u\bowtie v \}\downarrow
\end{array}\]
where, in the last line, $v$ is required to be proper.  Then we put:
\[(\async_{\PThread})_{P}(Q)  =  \bigcup_{u\in P, v\in Q} u \rlaction v\]
and $(\rsync_{\PThread})_P(Q) = (\async_{\PThread})_Q(P)$. With these definitions, 
$-^c$ is the extension of the  map  $\finalm \mapsto \halt_{\PThread}$ to $\Proc$.

In the converse direction one can consider adding missing algebraic operations to the language, for example adding $\cup$ and $\rsync$ via constructs $\orm{C}{D}$ and $\rfork{C}$. The latter construct is to the binary $\rsync$ as $\forkn$ is to the binary $\async$.
It generalizes $\yield$, which is equivalent to $\rfork \unit$.
Its operational semantics is given by the rule: 
\[\begin{array}{lcl@{\quad}l}
\transrulenssimple
        {\nt{\sigma,T,\avcxt{\rfork{C}}}}
        {\nt{\sigma,T.\avcxt{\unit},C}}
\end{array}\]

One may debate the programming usefulness of such additional constructs, but they do allow one to express the equations used for the algebraic characterizations at the level of commands. For example, the equation 
$P \rlaction \delay(x) =  \delay(P \bowtie x)$ 
becomes:
\[
\begin{array}{l}
(\fork{C});\yield;D\\ \qquad =\\ 
\yield; \orm{((\fork{C});D}{(\rfork{C});D)}
\end{array}\]

\subsection{Dendriform Algebras and Modules}

We have found it useful to employ various forms of shuffle: sometimes we shuffle two things of the same kind with each other, e.g.,  two pure transition sequences with each other; and sometimes we shuffle two things of different kinds with each other, e.g., a pure transition sequence with a transition sequence.

We have further found it useful to break down such shuffles into left and right shuffles, e.g., in the case of the left and right shuffles of asynchronous processes with processes; indeed we employ a uniform notation, writing $\llaction, \rlaction$, and 
$\bowtie$ for left shuffles, right shuffles, and (ordinary) shuffles, respectively. 
Our algebraic account of threads has further involved a number of equations concerning the interaction of these shuffle operations with each other and with other operations.

Shuffle operations and their algebra have been studied in a variety of settings. In particular, Loday's dendriform algebras~\cite{Lod01,FG08} provide a wide-ranging general notion of left and right shuffling of two things of the same kind with each other. Foissy's dendriform $A$-modules~\cite{Foi07} provide the corresponding notion of action: left or right shuffling a thing of one kind with a thing of another kind. We next relate our treatment to these general concepts, thereby placing our various shuffle operations and our equations for them in a standard algebraic context.

Let $R$ be a given commutative semiring (with no requirement for a $0$ or a $1$). 
Then a {\em dendriform dialgebra}  is an $R$-module $A$ equipped with two binary bilinear operations 
$\llaction$ and $\rlaction$ such that, for all $x,y,z \in A$:
\[\begin{array}{lcl}
(x \llaction y) \llaction z & = & x \llaction (y \bowtie z)\\
 x \rlaction (y \rlaction z) & = & (x \bowtie y) \rlaction z\\
(x \rlaction y) \llaction z & = & x \rlaction (y \llaction z)
\end{array}\]
where $x \bowtie y \eqdef x \llaction y + y \rlaction x$; it is \emph{commutative} if 
$x \llaction y = y\rlaction x$ always holds. Then $(A, \bowtie)$ is a semigroup in the category of $R$-modules, equivalently
$\bowtie$ is an associative bilinear operation;  it is commutative if the dialgebra is.

 Given a dendriform algebra $A$, a {\em dendriform $A$-module} is an $R$-module $M$ equipped with two binary bilinear operations 
$\llaction,\rlaction: A \times M \rightarrow M$ such that, for all $a,b \in A$ and $x \in M$:
\[\begin{array}{lcl}
(a \llaction b) \llaction x & = & a \llaction (b \bowtie x)\\
a \rlaction (b \rlaction  x) & = & (a \bowtie b) \rlaction x\\
(a \rlaction b) \llaction x & = & a \rlaction (b \llaction x)
\end{array}\]
where $\bowtie: A \times M \rightarrow M$ is given by: $a \bowtie x = a \llaction x + a \rlaction x$.
Then $\;\bowtie: A \times M \rightarrow M$   is a bilinear action of $(A,\bowtie)$ on $M$.

In all our examples we take  $R$ to be the natural two-element semiring over $\mathbb{B}$; join semilattices with a zero form $\mathbb{B}$-modules (setting $\btrue \,x = x$ and $\bfalse\, x  = 0$). 
 As a first example, 
consider the $\mathbb{B}$-module of the collection of all languages, i.e., all sets of strings over a given alphabet, not containing $\varepsilon$. This is 
a commutative dialgebra, taking $\llaction$ to be the left shuffle operation, and $\rlaction$ to be the right one; $\bowtie$ is then  the usual shuffle operation.

The semilattice of asynchronous processes $\PThread$ forms a commutative dendriform $\mathbb{B}$-algebra, setting:
\[P \llaction_{\PThread} Q = (\rsync_{\PThread})_P(Q) \quad\quad\quad P \rlaction_{\PThread} Q = (\async_{\PThread})_P(Q)\]
One then has that  $\QProc$ forms 
 a dendriform $\PThread$-module, setting:
\[P \llaction_{\scriptsize\QProc} I = (\rsync_{\Proc})_P(I) \quad\quad\quad P \rlaction_{\scriptsize\QProc} I =  (\async_{\Proc})_P(I)\]
It follows that  $\Proc$ also forms 
a dendriform $\PThread$-module, using the definitions of the left and right shuffling  given in Corollary~\ref{Proc-algebra}.

Algebraically, the first group of equations for $L_{\Spawn}$ state the bilinearity of the two module operations. The second group contains the second of the three module equations. 
The equation 
\[a \rlaction (b \llaction x)  = b \llaction (a \bowtie x)\]
generalizing one considered above, holds in any module over a commutative dendriform algebra.  To account for the other two module equations algebraically one would need an algebraic treatment of the dendriform algebra operations on $\PThread$. These operations are effect deconstructors rather than effect  constructors. An account of unary deconstructors has been given in~\cite{PP09}, but a satisfactory treatment of binary ones remains to be found; we therefore leave further algebraic treatment to future work.

\section{Conclusion}
\label{sec:conclusion}

A priori, the properties and the semantics of threads in general, and
of cooperative threads in particular, may not appear obvious. In our
opinion, a huge body of incorrect multithreaded software and a
relatively small literature both support this point of view.  With the
belief that mathematical foundations could prove beneficial, the
main technical goal of our work is to define and elucidate the
semantics of threads. For instance, semantics can serve for validating
reasoning principles; our work is only a preliminary, but encouraging,
step in this respect.

Our initial motivation was partly practical---we wanted to understand
and further the AME programming model and similar ones. We also saw an
opportunity to leverage developments in trace-based denotational
semantics and in the algebraic theory of effects, and to extend their
applicability to threads. As our results demonstrate, the convergence
of these three lines of work proved interesting and fruitful.

We focus on a particular small language with constructs for threads.
Several possible extensions may be considered.
These include constructs for parallel composition, nondeterministic
choice, higher-order functions, and thread-joining. 
More speculatively, they also include generalized yields, 
of the kind that
arise in the algebraic theory of effects, as discussed
in Section~\ref{sec:algebra}. 
Importantly, our monadic treatment of
threads indicates how to add higher-order functions to the semantics. 

Our results mostly carry over to these extensions.  
In some cases, small changes or restrictions are required. In
particular, the full-abstraction proof with nondeterministic choice
would use fresh variables; the one for higher-order functions might
require standard limitations on the order of functions, cf.~\cite{LICS::Jeffrey1995}. Thus, our
approach seems to be robust, and indeed---as in the case of
higher-order functions---helpful in accounting for a range of language
features. Further, our algebraic analysis of the thread monad links it to the broader theme of the algebraic treatment of effects. In that regard, as the discussion after Theorem~\ref{processes} indicates, there is clearly still further understanding to be gained.
 
Another possible direction for further work is the exploration of
alternative semantics.  For instance, we could 
switch from the ``may''
semantics that we study to ``must'' semantics.  We could 
also define
alternative notions of observation. As suggested in
Section~\ref{sec:fullabstractionII}, some of the coarser notions of
observation might 
require closure conditions, such as closure under
suitable forms of 
stuttering and under mumbling.  
These may correspond to suitable axioms on the suspension operator ${\delay}$, as alluded to in~\cite{Plo06}: we conjecture that stuttering corresponds to  ${\delay}({\delay}(x)) \leq \delay(x)$ 
and that mumbling corresponds to ${\delay}(x) \geq x$.  

It would also be interesting to consider finer notions of
observation that distinguish blocking from divergence. To this end we could  
add
constructs such as {\tt orElse}~\cite{harris:composable} and, in the semantics, treat blocking as a kind of exception. Finally, we
could 
revisit lower-level semantics with explicit optimistic
concurrency and roll-backs, of the kind employed in the implementation
of AME.

\section*{Acknowledgements}

We are grateful to Mart\'\i n Escard\'o and Martin Hyland for their helpful comments and suggestions.

\end{document}